\documentclass{article}

\usepackage{microtype}
\usepackage{graphicx}
\usepackage{subcaption}
\usepackage{subfiles}
\usepackage{multirow}
\usepackage{adjustbox}
\usepackage{float}
\usepackage{booktabs} 

\usepackage{hyperref}

\newcommand{\best}[1]{\textcolor{red}{\textbf{{#1}}}}

\usepackage[preprint]{icml2026}

\usepackage{amsmath}
\usepackage{amssymb}
\usepackage{mathtools}
\usepackage{amsthm}

\usepackage[capitalize,noabbrev]{cleveref}

\theoremstyle{plain}

\theoremstyle{definition}

\theoremstyle{remark}

\usepackage[textsize=tiny]{todonotes}

\icmltitlerunning{SWIFT: Spatio-temporal Wavelet Integrated Forecasting Framework for Workload Traces}

\begin{document}

\twocolumn[
  \icmltitle{SWIFT: Spatio-temporal Wavelet Integrated Forecasting \\Framework for Workload Traces}




  \begin{icmlauthorlist}
    \icmlauthor{Zeyuan Ding}{sjtu}
    \icmlauthor{Lingfeng Zheng}{snnu}
    \icmlauthor{Dian Ding}{sjtu}
    \icmlauthor{Guangtao Xue}{sjtu}
  \end{icmlauthorlist}

  \icmlaffiliation{sjtu}{Shanghai Jiao Tong University, Shanghai, China}
  \icmlaffiliation{snnu}{Shaanxi Normal University, Xi'an, China}

  \icmlcorrespondingauthor{Dian Ding}{dingdian94@sjtu.edu.cn}
  \icmlcorrespondingauthor{Guangtao Xue}{gt xue@sjtu.edu.cn}

  \icmlkeywords{Machine Learning, ICML}

  \vskip 0.3in
]



\printAffiliationsAndNotice{}  

\begin{abstract}
  Accurate cloud workload forecasting is pivotal for efficient resource management but remains challenging as workloads are highly volatile and prone to sudden bursts. Although wavelets preserve temporal locality, rigid fixed bases struggle with complex patterns and isolated processing neglects critical spatial dependencies. To address this, we propose SWIFT, a pure convolutional framework designed for high-efficiency workload forecasting. We introduce a Learnable Cascaded Wavelet Path that reformulates the traditional fixed wavelet bases into adaptive convolutional operators, enabling precise, data-driven feature peeling. Complementing this, our Multivariate Interaction Module sequentially models inter-variable spatial and intra-variable feature interactions to stabilize and refine noisy workload states. Extensive experiments demonstrate that SWIFT achieves SOTA accuracy with linear $O(L)$ complexity, reducing prediction error by up to 31.04\% while cutting latency by 79.74\%.
\end{abstract}

\section{Introduction}
\label{section:1}
\begin{figure}[t]
    \centering
    \includegraphics[width=0.95\linewidth]{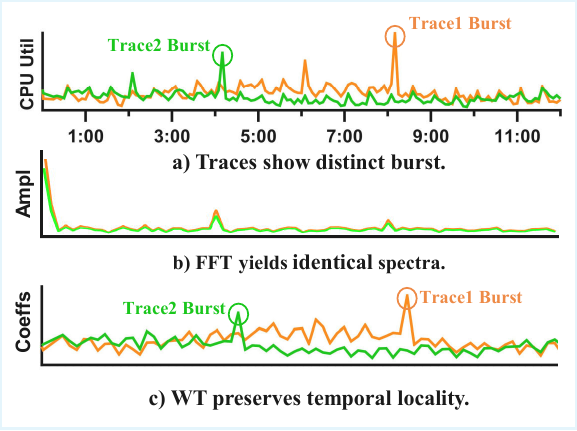} 
    
    \caption{\textbf{Motivation Analysis.} 
    (a) \textbf{Workload Characteristics:} Real-world cloud traces exhibit sudden bursts at distinct timestamps interlaced with multi-scale periodicities.
    (b) \textbf{FFT Inferiority:} Due to global integration, FFT blindly aggregates these distinct signals into indistinguishable spectra, effectively causing a loss of temporal locality.
    (c) \textbf{WT Superiority:} Unlike FFT, Wavelet Transform preserves temporal locality, enabling the precise distinction of burst timings and restoring their unique signatures}
    \label{fig:motivation}
    \vspace{-0.2in}
\end{figure}

\par Modern cloud environments leverage the synergy between containerized microservices to provide applications~\cite{1-1, 1-2}. To reconcile Quality of Service (QoS) with optimal resource utilization, cloud providers use workload forecasting to drive proactive resource management strategies, such as scaling~\cite{1-3, 1-4} and scheduling~\cite{1-5, 1-6}.

\par However, in large-scale cloud environments, achieving high-performance workload prediction faces two primary hurdles: 

\textbf{(1) Complex Workload Patterns:} Temporally, real-world cloud traces are highly volatile and prone to sudden bursts. As illustrated in Figure~\ref{fig:motivation}(a), distinct workloads often exhibit sharp, localized fluctuations occurring at completely different timestamps. Furthermore, these traces contain interlaced multi-scale periodicities (e.g., daily and hourly patterns) compounded with intense non-stationarity~\cite{zhao2024tfegru}. Spatially, microservice dependencies induce strong coupling and synchronized fluctuations across workload traces~\cite{ding2022coin}. Therefore, cloud workloads exhibit complex patterns that are difficult to capture.

\par Prior works~\cite{wu2022timesnet, frets, fredformer} primarily utilize the Fast Fourier Transform (FFT) to decouple these features. However, FFT's global basis functions cause a loss of temporal locality~\cite{mallat1999wavelet}. As evident in Figure~\ref{fig:motivation}(b), distinct workload bursts yield indistinguishable spectra. This limitation prevents the model from distinguishing when a burst occurs. Consequently, the predictor reacts untimely to sudden changes, predicting a spike only after it has already occurred, which defeats the purpose of proactive scaling.

\par To address this, recent studies have explored the Wavelet Transform (WT)~\cite{liao2020multiple,MODWT-LSTM,wdformer}. As shown in Figure~\ref{fig:motivation}(c), WT effectively captures time-frequency details, maintaining distinct signatures unlike FFT. However, existing methods typically rely on rigid decomposition schemes that struggle with intricate multi-scale periodicities and overlook spatial dependencies in coupled workloads, resulting in suboptimal performance.

\textbf{(2) Extreme Efficiency Requirements:} Platforms like ByteDance must process over 100,000 concurrent prediction tasks per hour~\cite{fremer}, demanding high precision with minimal computational overhead.
\par However, high-performance models often incur prohibitive costs. Transformer-based models~\cite{crossformer,informer,autoformer} suffer from quadratic $O(L^2)$ complexity, causing severe inference latency and memory exhaustion on long sequences. Similarly, conventional wavelet-based methods often rely on rigid, serial decomposition schemes that incur significant computational overhead due to unparallelizable implementations.
\par To mitigate these costs, Temporal Convolutional Networks~\cite{moderntcn, micn, group} offer a more efficient alternative with linear complexity. Yet, these models predominantly operate in the time domain and often neglect spatial correlations. This limitation prevents them from capturing complex, coupled workload patterns, leading to suboptimal accuracy.

\par To bridge the gap between precise pattern capture and deployment efficiency, we propose a novel paradigm: synergizing the multi-resolution analysis of Wavelet Transform with the efficiency of modern convolutional architectures. However, effectively integrating these technologies faces three fundamental technical challenges:

\textbf{Challenge 1: How to achieve effective multi-scale decoupling using WT?} 
    Standard wavelet schemes utilize shallow, fixed decompositions that fail to separate the intricate periodic patterns of cloud workloads. 

\textbf{Challenge 2: How to capture spatial correlations while leveraging WT?} 
    The Wavelet Transform is inherently designed for univariate signal analysis. However, cloud workloads exhibit strong spatial dependencies due to microservice coupling.

\textbf{Challenge 3: How to maximize efficiency and parallelism in WT implementations?} 
    Traditional wavelet algorithms are inherently sequential, which severely limits parallelization on modern GPUs. This creates a computational bottleneck for high-throughput tasks.

\par To address the above challenges, we propose \textbf{SWIFT}, a \textbf{S}patio-temporal \textbf{W}avelet \textbf{I}ntegrated \textbf{F}orecasting Framework for Workload \textbf{T}races. Specifically, we design: 
\par \textbf{(1) Learnable Cascaded Wavelet Path:} To address shallow decomposition, we design a hierarchical architecture that performs level-by-level feature peeling using learnable weights. This adaptive strategy overcomes the rigidity of fixed bases, enabling the precise decoupling of intricate multi-scale patterns with minimal parameter overhead.
\par \textbf{(2) Multivariate Interaction Module:} To complement the univariate wavelet path, MIM captures spatial correlations via a sequential architecture. It prioritizes inter-variable interaction to stabilize workload states before facilitating intra-variable feature interaction, bridging the gap between isolated decomposition and system-wide coupling.
\par \textbf{(3) Hardware-Friendly Pure Convolutional Architecture:} To ensure deployment efficiency, we replace CPU-based recursive calls with GPU-native depthwise convolutions. This design maintains linear $O(L)$ complexity while eliminating CPU-GPU data transfer bottlenecks, thereby achieving superior inference throughput.

\par Extensive experiments on real-world cloud workload datasets demonstrate that SWIFT significantly outperforms sota models. Specifically, it reduces prediction error by up to \textbf{31.04\%} in workload forecasting tasks and \textbf{20.05\%} in generalization scenarios, while cutting inference latency by up to \textbf{79.74\%}.
\section{Related Work}
\label{section:2}
\subsection{Fourier-based Forecasting Methods}
\par Since time series data frequently exhibit strong periodicity and complex patterns, recent research has increasingly shifted toward frequency-domain modeling to capture global features. For instance, TimesNet~\cite{wu2022timesnet} identifies dominant periods via FFT to reshape 1D signals into 2D tensors, enabling 2D kernels to capture complex variation patterns. Fredformer~\cite{fredformer} mitigates frequency bias by employing local normalization and frequency patching to balance learning across all spectral components with linear computational efficiency. FreTS~\cite{frets} adopts a minimalist approach by using MLPs to learn directly from the real and imaginary components of frequency spectra. Fremer~\cite{fremer} employs complex-valued spectral attention to achieve deep decoupling of complex workload patterns within the frequency domain. However, these global basis-based methods often overlook signal non-stationarity and lose temporal locality, leading to phase lags during abrupt workload shifts.

\subsection{Wavelet-based Forecasting Methods}
\par Wavelet-based time series forecasting methods leverage multi-resolution analysis to effectively extract local features in both the time and frequency domains. By concatenating the original data with multiple reconstructed sequences, MWCNN~\cite{liao2020multiple} effectively captures potential periodic patterns and achieves high robustness against input disturbances. WDformer~\cite{wdformer} integrates wavelet transforms directly into a Transformer architecture and employs a differential attention mechanism to enhance the capture of core trends and periodic structures. The MODWT-LSTM~\cite{MODWT-LSTM} leverages Maximum Overlap Discrete Wavelet Transform to overcome the sample size and translation sensitivity constraints of traditional DWT, achieving significantly improved non-linear forecasting accuracy through Hyperband-optimized LSTM modeling on decomposed sub-series. However, these rigid schemes fail to fully decouple complex periodicities and incur high overhead due to unparallelizable implementations. Furthermore, they overlook spatial dependencies by isolating coupled workloads, leading to suboptimal performance.

\subsection{Tranformer-based Forecasting Methods}
\par Recently, Transformers have achieved breakthroughs in various domains, spawning diverse paradigms for time series forecasting. Specifically, Informer~\cite{informer} and TimeXer~\cite{timexer} focus on capturing long-range temporal dependencies. Decomposition-based models like Autoformer~\cite{autoformer} and FEDformer~\cite{zhou2022fedformer} enhance robustness by decoupling series into trend, periodic, and residual components. Meanwhile, Crossformer~\cite{crossformer} and iTransformer~\cite{itransformer} prioritize inter-variable spatial correlations. Nervertheless, the quadratic computational complexity of self-attention mechanisms leads to prohibitive resource consumption and latency in massive-scale production platforms.

\subsection{Convolution-based Forecasting Methods}
\par Recent studies have revitalized Convolutional Neural Networks (CNNs) for time series forecasting. MICN~\cite{micn} proposes a multi-scale isometric convolution framework with a dual-branch structure, utilizing down-sampling 1D convolutions to extract local features and isometric convolutions to capture global correlations. Group~\cite{group} employs Inception-style convolutional layers  to extract multi-feature evolution patterns. ModernTCN~\cite{moderntcn} adapts the design principles of modern vision backbones by incorporating large kernel sizes and depth-wise separable convolutions, which models long-range temporal patterns while maintaining high computational efficiency. However, despite superior efficiency over attention-based models, these methods primarily focus on the time domain and often neglect spatial correlations. This limitation hinders capturing complex coupling effects in distributed workloads, leading to suboptimal prediction accuracy.
\section{Preliminary}
\label{sec:preliminary}

In this section, we analyze the limitations of the Fast Fourier Transform in modeling non-stationary cloud workloads and mathematically justify the selection of the Wavelet Transform as the theoretical foundation of SWIFT.

\subsection{Limitations of Fast Fourier Transform}

\par The Fast Fourier Transform projects a time series $X \in \mathbb{R}^L$ onto a set of orthogonal sinusoidal basis functions. For a frequency $k$, the spectral component $X_k$ is defined as:
\begin{equation}
    X_k = \sum_{n=0}^{L-1} x_n \cdot e^{-i \frac{2\pi}{L} k n}
\end{equation}
The magnitude $|X_k|$ represents the global strength of frequency $k$. However, the summation operator $\sum_{n=0}^{L-1}$ integrates information over the entire time domain $[0, L-1]$. This global integration leads to the loss of temporal locality. 

\par As empirically evident in Figure 1(b), distinct workload traces with totally different temporal distributions can yield indistinguishable spectral signatures. Since $|X_k|$ aggregates the energy over the entire sequence, the FFT spectrum inherently fails to preserve the temporal localization of workload fluctuations. Consequently, the model fails to differentiate between these patterns, leading to poor responsiveness.

\par Furthermore, approximating non-stationary abrupt changes with continuous sinusoids triggers the Gibbs phenomenon due to frequency truncation. The resulting spurious oscillations manifest as phase lags, causing the model to react sluggishly to sudden workload changes.

\subsection{Advantages of Wavelet Transform}
To address these issues, we adopt the Wavelet Transform, which decomposes signals using a family of functions $\psi_{a,b}(t)$ derived from a mother wavelet $\psi(t)$ via scaling ($a$) and translation ($b$):
\begin{equation}
    \psi_{a,b}(t) = \frac{1}{\sqrt{|a|}} \psi\left(\frac{t-b}{a}\right)
\end{equation}
Unlike the infinite-duration sinusoids in FFT, the wavelet function $\psi(t)$ has compact support (it is localized in time). The transform coefficient $W(a,b)$ measures the similarity between the signal and the wavelet at scale $a$ and location $b$:
\begin{equation}
    W(a,b) = \int_{-\infty}^{\infty} x(t) \psi_{a,b}^*(t) dt
\end{equation}
This formulation fundamentally enhances workload forecasting by offering simultaneous time-frequency localization. By utilizing the translation parameter $b$ to preserve temporal locality and the scale parameter $a$ for multi-resolution analysis, the model can explicitly distinguish localized stochastic bursts from global periodic trends while capturing high-frequency spikes with high precision. This intrinsic capability theoretically resolves the "indistinguishable spectra" problem and minimizes the phase lags and Gibbs artifacts typical of rigid Fourier basis functions, offering the potential for responsive forecasting.
\section{Methodology}
\label{section:3}
\begin{figure*}[t]
    \centering
    \includegraphics[width=0.95\linewidth]{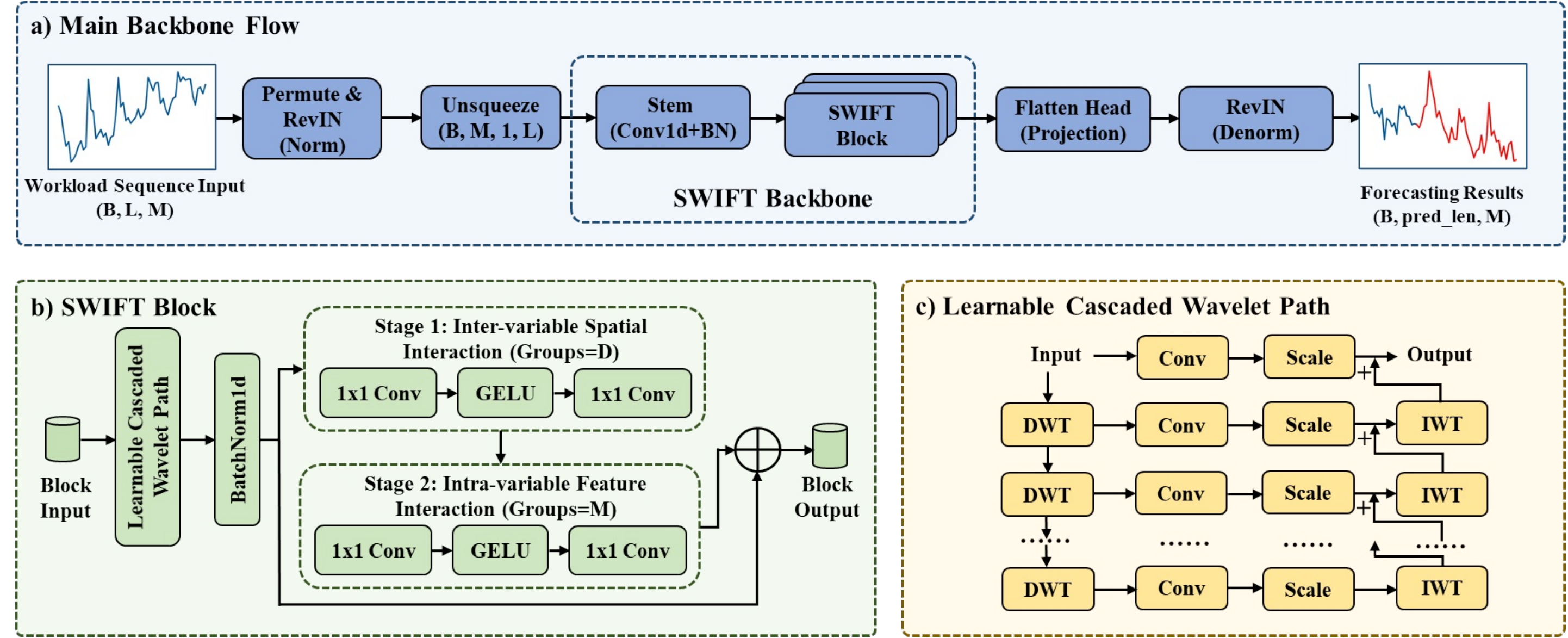}
    \caption{\textbf{Overview of the SWIFT Framework.} 
    \textbf{(a) Main Backbone Flow:} The workload sequence undergoes Reversible Instance Normalization (RevIN) before entering the stacked SWIFT Blocks. 
    \textbf{(b) SWIFT Block:} Each block integrates the \textbf{Learnable Cascaded Wavelet Path} for temporal modeling, followed by a two-stage \textbf{Multivariate Interaction Module (MIM)} for spatial-feature fusion. 
    \textbf{(c) Learnable Cascaded Wavelet Path:} This module recursively decomposes workloads via convolutional DWT (left-to-right). At each level, learnable convolutions (Conv) refine coefficients to correct morphological artifacts, while scaling factors (Scale) filter spectral noise. The signal is finally reconstructed via IWT (right-to-left) to yield high-fidelity representations.}
    \label{fig:SWIFT}
\end{figure*}

\par \par In this section, we detail \textbf{SWIFT}, the Spatio-temporal Wavelet Integrated Forecasting Framework for Workload Traces. As illustrated in Figure~\ref{fig:SWIFT}(a), SWIFT adopts a pure convolutional backbone. To systematically address the challenges of complex workload pattern adaptability and computational efficiency, we introduce our core contributions in the following subsections: the Learnable Cascaded Wavelet Path for temporal modeling, and the Multivariate Interaction Module for spatial modeling.

\subsection{Learnable Cascaded Wavelet Path}

Standard wavelet implementations typically rely on CPU-centric recursive algorithms, which hinder parallelization on GPUs. Furthermore, fixed basis functions struggle to adapt to the non-stationary drifts in cloud workloads. To overcome these limitations, we propose the \textbf{Learnable Cascaded Wavelet Path} (Figure~\ref{fig:SWIFT}(c)), a module that re-engineers the discrete wavelet transform (DWT) into a fully parallelizable, deep, and data-driven architecture.

\paragraph{Convolutional Re-engineering of DWT.} 
Instead of relying on standard recursive filter banks, we reformulate the DWT and Inverse DWT (IWT) as parallelizable convolution operations. 
Inspired by~\cite{finder2024wavelet}, for a workload sequence $X$, the decomposition into approximation ($X_{LL}$) and detail ($X_H$) components is executed via a fixed-weight depth-wise convolution with a stride of 2:
\begin{equation}
    [X_{LL}, X_H] = \text{Conv1D}_{\text{stride}=2}([g_L, g_H], X)
\end{equation}
where $g_L$ and $g_H$ are the low-pass and high-pass filters (e.g., Haar), respectively. This reformulation allows the entire wavelet transformation to be executed natively on Tensor Cores, eliminating CPU-GPU context switching and ensuring high throughput.

\paragraph{Cascaded Feature Peeling.} 
To decouple intricate multi-scale periodicities, we extend the shallow wavelet decomposition into a deep cascaded structure. 
As illustrated in the forward path of Figure~\ref{fig:SWIFT}(c), unlike independent multi-level decompositions, we recursively feed the low-frequency component $X_{LL}^{(i)}$ from the current level as the input to the next level $i+1$.
This creates a "feature peeling" effect: as the decomposition depth increases (moving rightwards in the figure), the receptive field expands exponentially. This allows the model to progressively isolate high-frequency stochastic bursts in shallower layers while preserving long-range trends in deeper layers.

\paragraph{Learnable Coefficient Rectification.} 
To overcome the rigidity of fixed basis functions, we introduce a two-step learnable mechanism at each decomposition level:

\textit{(1) Morphological Feature Correction:} As shown in the \textbf{Conv} block of Figure~\ref{fig:SWIFT}(c), we process the coefficients through a learnable convolution layer $W^{(i)}$. This transforms static wavelet coefficients into adaptable representations:
\begin{equation}
    \tilde{Y}^{(i)} = \text{Conv}(W^{(i)}, \text{Concat}[X_{LL}^{(i)}, X_H^{(i)}])
\end{equation}
Physically, this acts as a morphological corrector, smoothing out artifacts introduced by the fixed filters and aligning features with actual workload shapes.

\textit{(2) Adaptive Spectral Filtering:} We then apply a channel-wise scaling vector (labeled as \textbf{Scale}) to dynamically re-weight frequency components:
\begin{equation}
    Y^{(i)} = S^{(i)} \odot \tilde{Y}^{(i)}
\end{equation}
Functionally, this acts as a soft spectral filter, suppressing channels dominated by stochastic noise while amplifying informative periodic signals.

\paragraph{Aggregated Reconstruction.} Finally, the refined features are progressively reconstructed via the IWT branch (Figure~\ref{fig:SWIFT}c, bottom path). The output $Z^{(i)}$ is fused with the reconstruction from the deeper level $Z^{(i+1)}$ in a bottom-up manner:
\begin{equation}
    Z^{(i)} = \text{IWT}(Y_{LL}^{(i)} + Z^{(i+1)}, Y_H^{(i)})
\end{equation}
This ensures that the final representation encapsulates refined information from all frequency bands.

\subsection{Multivariate Interaction Module}

While the Wavelet Path excels at decoupling multi-scale temporal patterns, it processes each sequence in isolation. However, cloud systems exhibit strong spatial coupling (e.g., microservice dependencies). To capture these correlations without the quadratic complexity of Attention, we propose the \textbf{Multivariate Interaction Module (MIM)}, located within the SWIFT Block (Figure~\ref{fig:SWIFT}(b)). MIM employs a two-stage design based purely on Grouped Convolutions.

\paragraph{Stage 1: Inter-variable Spatial Interaction.} 
The first stage bridges isolated variables to leverage system-wide context. We utilize grouped $1\times1$ convolutions where the number of groups equals the feature dimension $D$:
\begin{equation}
    Z_{spatial} = \text{GroupConv}_{g=D}(\text{Permute}(Z_{in}))
\end{equation}
By mixing information across the variable dimension $M$, this stage captures global spatial dependencies. Since raw traces are noisy, interacting with correlated neighbors helps stabilize the workload state before detailed feature extraction.

\paragraph{Stage 2: Intra-variable Feature Interaction.} 
The second stage focuses on dense information exchange within each variable's feature embedding. Here, we apply grouped convolutions with groups set to the variable dimension $M$:
\begin{equation}
    Z_{out} = \text{GroupConv}_{g=M}(Z_{spatial})
\end{equation}
This enables non-linear mapping of the feature channels for each variable independently. By decoupling spatial mixing and feature mixing, MIM efficiently captures the comprehensive state of the cloud system with linear complexity $O(L)$, perfectly complementing the temporal focus of the wavelet path.
\section{Experiments}
\label{section:4}
\subsection{Experimental Setup}
\paragraph{Datasets.}
Our experiments are primarily conducted on four real-world cloud workload datasets: the Alibaba cluster-trace-microservices-v2021~\cite{5-1} (\textbf{Ali2021}), Alibaba cluster-trace-microservices-v2022~\cite{5-2} (\textbf{Ali2022}), \textbf{Fisher}~\cite{5-3}, and Google 2011 cluster data~\cite{5-4} (\textbf{Google}). Additionally, we include the ETT benchmarks~\cite{informer} to verify universality. Dataset details are provided in Appendix~\ref{sec: append-datasets}.

\paragraph{Baselines.}
To comprehensively evaluate the performance of SWIFT, we extensively include the latest and advanced forecasting models as baselines, covering two mainstream technical paradigms. Specifically, we compare SWIFT against \textbf{Transformer-based models}, including TimeXer~\cite{timexer}, Fremer~\cite{fremer}, and WDFormer~\cite{wdformer}; and \textbf{CNN-based models}, including TimesNet~\cite{wu2022timesnet}, ModernTCN~\cite{moderntcn}, and MICN~\cite{micn}. Notably, among these baselines, \textbf{Fremer} and \textbf{TimesNet} utilize FFT for frequency domain analysis, while \textbf{WDFormer} incorporates wavelet decomposition. This diverse selection ensures a rigorous validation against sota methods with different inductive biases.

\paragraph{Evaluation Metrics.}
\par We calculate the RMSE, MAE, MSE, and SMAPE as evaluation metrics. The detailed definitions and formulas for these metrics are provided in the Appendix \ref{sec: append-metric}. 

\subsection{Long-term Workload Forecasting}

\begin{table*}[t!]
    \centering
    \caption{\textbf{Long-Term Workload Forecasting Results.} We report the average RMSE and MAE across all four prediction horizons ($T \in \{64, 128, 192, 256\}$). \best{Red} indicates the best result. See Table~\ref{long-term-append} in Appendix for full results.}
    \label{long-term}
    
    \renewcommand{\arraystretch}{1.1} 
    \setlength{\tabcolsep}{3pt}        

    \begin{adjustbox}{width=\linewidth}
    \begin{tabular}{l|cc|cc|cc|cc|cc|cc|cc}
        \toprule
        Models & 
        \multicolumn{2}{c|}{\textbf{SWIFT}} & 
        \multicolumn{2}{c}{TimesNet} &    
        \multicolumn{2}{c}{TimeXer} &       
        \multicolumn{2}{c}{Fremer} &       
        \multicolumn{2}{c}{WDFormer} &        
        \multicolumn{2}{c}{ModernTCN} &    
        \multicolumn{2}{c}{MICN} \\    
        
        \cmidrule(r){1-1} \cmidrule(lr){2-3} \cmidrule(lr){4-5} \cmidrule(lr){6-7} \cmidrule(lr){8-9} \cmidrule(lr){10-11} \cmidrule(lr){12-13} \cmidrule(l){14-15}
        
        Metric & RMSE & MAE & RMSE & MAE & RMSE & MAE & RMSE & MAE & RMSE & MAE & RMSE & MAE & RMSE & MAE \\
        \midrule

        Ali2021 & 0.1084 & \best{0.0793} & 0.1213 & 0.0901 & 0.1103 & 0.0805 & 0.1104 & 0.0805 & 0.1101 & 0.0805 & 0.1118 & 0.0814 & \best{0.1080} & 0.0800 \\ \midrule
        
        Ali2022 & \best{0.0479} & \best{0.0321} & 0.0901 & 0.0347 & 0.0501 & 0.0363 & 0.0482 & 0.0324 & 0.0487 & 0.0328 & 0.0483 & 0.0324 & 0.0490 & 0.0330 \\ \midrule
        
        Fisher  & \best{0.0737} & \best{0.0495} & 0.0772 & 0.0538 & 0.0752 & 0.0503 & 0.0757 & 0.0503 & 0.0781 & 0.0535 & 0.0749 & 0.0749 & 0.0754 & 0.0503 \\ \midrule
        
        Google  & \best{0.0610} & \best{0.0400} & 0.0653 & 0.0411 & 0.0618 & 0.0411 & 0.0612 & 0.0403 & 0.0636 & 0.0427 & 0.0618 & 0.0405 & 0.0665 & 0.0461 \\
        
        \bottomrule
    \end{tabular}
    \end{adjustbox}
\end{table*}

\paragraph{Setup.} 
For all datasets, we set the input look-back window $T=512$ and evaluate prediction horizons $H \in \{64, 128, 192, 256\}$. The reported results represent the average RMSE and MAE across these four horizons.

\paragraph{Results.} 
Our main results for long-term workload forecasting are summarized in Table~\ref{long-term}. SWIFT demonstrates excellent performance across diverse cloud workload domains, achieving the sota in \textbf{7 out of 8} cases. Compared to advanced Transformer-based baselines, SWIFT reduces the prediction error by up to \textbf{11.58\%}. SWIFT also outperforms recent CNN-based baselines with a maximum error reduction of \textbf{13.14\%}. These results highlight SWIFT's capability in effectively capturing the long-range temporal dependencies and spatial correlations among workloads.

\subsection{Short-term Workload Forecasting}

\begin{table}[t]
    \centering
    \caption{\textbf{Short-Term Workload Forecasting Results.} \best{Red} indicates the best result.}
    \label{short-term}

    \renewcommand{\arraystretch}{1.2} 
    \setlength{\tabcolsep}{3pt} 

    \begin{adjustbox}{width=\linewidth}
    \begin{tabular}{cc|ccccccc}
        \toprule
        \multicolumn{2}{c|}{\textbf{Models}} & 
        \textbf{SWIFT} & TimesNet & TimeXer & Fremer & WDFormer & ModernTCN & MICN \\
        
        \midrule

        \multirow{3}{*}{\rotatebox{90}{Ali2021}} 
        & RMSE  & \best{0.0749} & 0.0940 & 0.0754 & 0.0879 & 0.0752 & 0.0760 & 0.0935 \\
        & MAE   & \best{0.0489} & 0.0671 & 0.0501 & 0.0615 & 0.0497 & 0.0502 & 0.0671 \\
        & SMAPE & \best{0.0841} & 0.1120 & 0.0858 & 0.1082 & 0.0857 & 0.0863 & 0.1190 \\ 
        \midrule

        \multirow{3}{*}{\rotatebox{90}{Ali2022}} 
        & RMSE  & \best{0.0438} & 0.0461 & 0.0450 & 0.0452 & 0.0450 & 0.0453 & 0.0449 \\
        & MAE   & \best{0.0280} & 0.0303 & 0.0294 & 0.0295 & 0.0290 & 0.0295 & 0.0302 \\
        & SMAPE & \best{0.1381} & 0.1447 & 0.1409 & 0.1408 & 0.1382 & 0.1392 & 0.1436 \\ 
        \midrule

        \multirow{3}{*}{\rotatebox{90}{Fisher}} 
        & RMSE  & \best{0.0266} & 0.0304 & 0.0279 & 0.0353 & 0.0280 & 0.0279 & 0.0297 \\
        & MAE   & \best{0.0162} & 0.0198 & 0.0174 & 0.0229 & 0.0173 & 0.0177 & 0.0191 \\
        & SMAPE & \best{0.0376} & 0.0442 & 0.0405 & 0.0546 & 0.0401 & 0.0397 & 0.0668 \\ 
        \midrule

        \multirow{3}{*}{\rotatebox{90}{Google}} 
        & RMSE  & \best{0.0580} & 0.0596 & 0.0587 & 0.0593 & 0.0589 & 0.0583 & 0.0590 \\
        & MAE   & \best{0.0372} & 0.0388 & 0.0384 & 0.0383 & 0.0385 & 0.0377 & 0.0381 \\
        & SMAPE & \best{0.1770} & 0.1846 & 0.1847 & 0.1831 & 0.1843 & 0.1782 & 0.1981 \\ 
        
        \bottomrule
    \end{tabular}
    \end{adjustbox}
\end{table}

\paragraph{Setup.} To assess the model's responsiveness to immediate workload fluctuations, we unify the setting across all datasets with an input length of $T=32$ and a prediction horizon of $H=8$. For this task, we additionally employ SMAPE alongside RMSE and MAE to capture relative prediction accuracy in low-load regions.

\paragraph{Results.}
The results are summarized in Table~\ref{short-term}. Short-term workload forecasting on these real-world cloud workload traces is challenging due to the extreme volatility and stochastic bursts inherent in user request patterns. Despite these difficulties, SWIFT consistently achieves sota performance across all datasets (\textbf{24.49\%} reduction on RMSE, \textbf{28.72\%} reduction on MAE, and \textbf{31.04\%} reduction on SMAPE), demonstrating its superior capability in capturing transient temporal dependencies and responsiveness to immediate workload fluctuations.

\subsection{Generalization Analysis}
\par To validate SWIFT's generalization capability, we conduct two sets of transfer experiments using a fixed look-back window of $L=512$ across horizons $T \in \{64, 128, 192, 256\}$.

\subsubsection{Intra-Dataset Generalization}
\begin{table*}[t!]
    \centering
    \caption{\textbf{Intra-Dataset Generalization Results.} Models are trained on the first 50\% of each dataset and evaluated on the remaining 50\%. Results are averaged across four horizons ($T \in \{64, 128, 192, 256\}$). \best{Red} indicates the best result. See Table~\ref{intra-data-append} in Appendix for full results.}
    \label{intra-data}
    
    \renewcommand{\arraystretch}{1.1} 
    \setlength{\tabcolsep}{3pt}        

    \begin{adjustbox}{width=\linewidth}
    \begin{tabular}{l|cc|cc|cc|cc|cc|cc|cc}
        \toprule
        Models & 
        \multicolumn{2}{c|}{\textbf{SWIFT}} & 
        \multicolumn{2}{c}{TimesNet} &    
        \multicolumn{2}{c}{TimeXer} &       
        \multicolumn{2}{c}{Fremer} &       
        \multicolumn{2}{c}{WDFormer} &        
        \multicolumn{2}{c}{ModernTCN} &    
        \multicolumn{2}{c}{MICN} \\    
        
        \cmidrule(r){1-1} \cmidrule(lr){2-3} \cmidrule(lr){4-5} \cmidrule(lr){6-7} \cmidrule(lr){8-9} \cmidrule(lr){10-11} \cmidrule(lr){12-13} \cmidrule(l){14-15}
        
        Metric & RMSE & MAE & RMSE & MAE & RMSE & MAE & RMSE & MAE & RMSE & MAE & RMSE & MAE & RMSE & MAE \\
        \midrule

        Ali2021 & \best{0.0696} & \best{0.0523} & 0.0751 & 0.0576 & 0.0726 & 0.0561 & 0.0721 & 0.0547 & 0.0769 & 0.0596 & 0.0700 & 0.0528 & 0.0746 & 0.0578 \\ \midrule
        
        Ali2022 & \best{0.0402} & \best{0.0288} & 0.0408 & 0.0298 & 0.0404 & 0.0292 & 0.0402 & 0.0291 & 0.0406 & 0.0296 & 0.0405 & 0.0296 & 0.0415 & 0.0302 \\ \midrule
        
        Fisher      & \best{0.0614} & \best{0.0431} & 0.0667 & 0.0473 & 0.0640 & 0.0451 & 0.0623 & 0.0438 & 0.0670 & 0.0480 & 0.0619 & 0.0436 & 0.0675 & 0.0472 \\ \midrule
        
        Google  & \best{0.0420} & \best{0.0261} & 0.0434 & 0.0277 & 0.0425 & 0.0263 & 0.0421 & 0.0264 & 0.0443 & 0.0281 & 0.0424 & 0.0265 & 0.0443 & 0.0286 \\
        
        \bottomrule
    \end{tabular}
    \end{adjustbox}
\end{table*}
\paragraph{Setup.} 
To evaluate the intra-dataset generalization capability, we randomly partition each dataset into two disjoint subsets of equal size (50\%/50\%). We train the model on the first half and directly test it on the second unseen half. This strict separation prevents information leakage, challenging the model to learn universal temporal patterns rather than memorizing sequence identities.

\paragraph{Results.}
Table~\ref{intra-data} summarizes the intra-dataset generalization performance, where SWIFT achieves the best results across all \textbf{8} cases. SWIFT demonstrates exceptional adaptability to unseen workload sequences, significantly outperforming competitive baselines. Notably, it reduces the prediction error by up to \textbf{12.16\%} compared to Transformer-based models and \textbf{9.49\%} compared to CNN-based models. This consistent superiority confirms that SWIFT successfully captures the universal temporal dynamics shared across the cloud environment, rather than merely memorizing specific sequence identities during training.

\subsubsection{Cross-Dataset Generalization}

\begin{table*}[t!]
    \centering
    \caption{\textbf{Cross-Dataset Generalization Results.} Models are trained solely on \textbf{Ali2022} and directly evaluated on the other datasets without fine-tuning. Results are averaged across four horizons ($T \in \{64, 128, 192, 256\}$). \best{Red} indicates the best result. See Table~\ref{cross-data-append} in Appendix for full results.}
    \label{cross-data}
    
    \renewcommand{\arraystretch}{1.1} 
    \setlength{\tabcolsep}{3pt}        

    \begin{adjustbox}{width=\linewidth}
    \begin{tabular}{l|cc|cc|cc|cc|cc|cc|cc}
        \toprule
        Models & 
        \multicolumn{2}{c|}{\textbf{SWIFT}} & 
        \multicolumn{2}{c}{TimesNet} &    
        \multicolumn{2}{c}{TimeXer} &       
        \multicolumn{2}{c}{Fremer} &       
        \multicolumn{2}{c}{WDFormer} &        
        \multicolumn{2}{c}{ModernTCN} &    
        \multicolumn{2}{c}{MICN} \\    
        
        \cmidrule(r){1-1} \cmidrule(lr){2-3} \cmidrule(lr){4-5} \cmidrule(lr){6-7} \cmidrule(lr){8-9} \cmidrule(lr){10-11} \cmidrule(lr){12-13} \cmidrule(l){14-15}
        
        Metric & RMSE & MAE & RMSE & MAE & RMSE & MAE & RMSE & MAE & RMSE & MAE & RMSE & MAE & RMSE & MAE \\
        \midrule

        Ali2021 & \best{0.0741} & \best{0.0550} & 0.0754 &  0.0563 &  0.0756 &  0.0561 &  0.0744 &  0.0552 &  0.0752 &  0.0560 &  0.0758 &  0.0567 &  0.0757 &  0.0569 \\ \midrule
        
        Fisher      & \best{0.0555} & \best{0.0380} &   0.0627 &  0.0429 &  0.0564 &  0.0388 &  0.0560 &  0.0386 &  0.0581 &  0.0405 &  0.0574 &  0.0397 &  0.0617 &  0.0433 \\  \midrule
        
        Google  & \best{0.0606} & \best{0.0416} & 0.0640 &  0.0452 &  0.0611 &  0.0428 &  0.0616 &  0.0418 &  0.0628 &  0.0440 &  0.0629 &  0.0438 &  0.0676 &  0.0500 \\ 
        
        \bottomrule
    \end{tabular}
    \end{adjustbox}
\end{table*}

\paragraph{Setup.} 
To evaluate the cross-dataset generalization capability, we train the model solely on Ali2022 and directly test it on Ali2021, Google, and Fisher without fine-tuning. This setting exposes the model to significant distribution shifts caused by varying sampling rates and resource usage patterns. Consequently, it rigorously tests the model's ability to capture platform-invariant workload dynamics.

\paragraph{Results.}
Table~\ref{cross-data} presents the zero-shot cross-dataset generalization results, where models trained on Ali2022 are directly deployed to distinct environments (Ali2021, Fisher, Google). Despite the severe distributional shifts arising from different infrastructure providers and temporal spans, SWIFT maintains its superiority, achieving sota results across all target domains. Quantitatively, SWIFT outperforms the Transformer-based model with a maximum error reduction of \textbf{6.12\%}. Compared to the CNN-based baseline, SWIFT achieves a significant improvement of up to \textbf{16.69\%}. This robustness confirms that SWIFT successfully captures universal multi-scale periodicities shared across cloud workloads, rather than overfitting to the specific statistical characteristics of the source domain.

\subsection{Efficiency-Effectiveness Analysis}

\begin{table}[t!]
    \centering
\caption{\textbf{Efficiency-Effectiveness Analysis.} We report the average RMSE, MAE, inference latency, and memory usage averaged across all four prediction horizons ($T \in \{64, 128, 192, 256\}$). See Table~\ref{efficiency-append} in Appendix for full results.}
    \label{efficiency-main}
    
    \renewcommand{\arraystretch}{1.15} 
    \setlength{\tabcolsep}{3pt}       

    \begin{adjustbox}{width=\linewidth}
    \begin{tabular}{cc|c|cccccc}
        \toprule
        \multicolumn{2}{c|}{\textbf{Models}} & 
        \textbf{SWIFT} & 
        TimesNet & TimeXer & Fremer & WDFormer & ModernTCN & MICN \\
        
        \midrule

        \multirow{4}{*}{\rotatebox{90}{Ali2021}}
        & RMSE & 0.0741 & 0.0754 & 0.0756 & 0.0744 & 0.0752 & 0.0758 & 0.0757 \\
        & MAE  & 0.0550 & 0.0563 & 0.0561 & 0.0552 & 0.0560 & 0.0567 & 0.0569 \\
        & Latency (ms) & 3.41 & 16.86 &  8.15 &  7.17 &  9.48 &  4.74 &  4.50 \\
        & Mem (MB)   & 136.82& 202.4 & 69.21 & 69.38 & 84.58 & 190.98 & 161.63 \\ 
        \midrule

        \multirow{4}{*}{\rotatebox{90}{Ali2022}}
        & RMSE & 0.0741 & 0.0754 & 0.0756 & 0.0744 & 0.0752 & 0.0758 & 0.0757 \\
        & MAE  & 0.0550 & 0.0563 & 0.0561 & 0.0552 & 0.0560 & 0.0567 & 0.0569 \\
        & Latency (ms) & 4.52 & 18.92 &  11.62 &  8.69 &  11.93 &  4.87 &  4.68 \\
        & Mem (MB)   & 2306.75 & 2152.54 & 1805.77 & 1785.32 & 1796.80 & 3629.20 & 3336.51 \\
        \midrule

        \multirow{4}{*}{\rotatebox{90}{Fisher}} 
        & RMSE & 0.0555 & 0.0627 & 0.0564 & 0.0560 & 0.0581 & 0.0574 & 0.0617 \\
        & MAE  & 0.0380 & 0.0429 & 0.0388 & 0.0386 & 0.0405 & 0.0397 & 0.0433 \\
        & Latency (ms) & 3.84 & 17.96 &  11.59 &  7.48 &  11.78 &  3.93 &  4.24 \\
        & Mem (MB)   & 188.29 & 440.21	& 92.39	& 83.60	& 98.31	& 251.09 & 189.17 \\ 
        \midrule

        \multirow{4}{*}{\rotatebox{90}{Google}} 
        & RMSE & 0.0606 & 0.0640 & 0.0611 & 0.0616 & 0.0628 & 0.0629 & 0.0676 \\
        & MAE  & 0.0416 & 0.0452 & 0.0428 & 0.0418 & 0.0440 & 0.0438 & 0.0500 \\
        & Latency (ms) & 3.71 &  17.32 &  10.85 &  7.46 &  10.83 &  3.72 & 4.43 \\ 
        & Mem (MB)   & 178.19 & 424.42& 115.48  & 106.61 & 121.32 & 273.91 & 191.54 \\
        
        \bottomrule
    \end{tabular}
    \end{adjustbox}
\end{table}

\paragraph{Setup.}
We evaluate the efficiency-effectiveness trade-off across four datasets. We report the average RMSE and MAE to measure effectiveness, alongside Inference Latency (ms/sample, with a batch size of 1) and GPU Memory Usage (MB) to quantify efficiency.

\paragraph{Results.} Table \ref{efficiency-main} highlights SWIFT's superior balance between efficiency and performance. By leveraging a hardware-friendly pure convolutional architecture that avoids attention bottlenecks, SWIFT maximizes GPU parallelization to achieve the \textbf{fastest inference speed} across all datasets. Specifically, SWIFT reduces inference latency by \textbf{66.82\%} compared to Transformer-based baselines and by \textbf{79.74\%} against CNN-based baselines. Furthermore, SWIFT secures the lowest prediction errors while remaining more memory-efficient than other CNN models. Although lightweight Transformers like Fremer utilize less memory via patching, they compromise on accuracy and latency. Consequently, SWIFT achieves an optimal trade-off, combining sota accuracy with the high throughput required for latency-sensitive cloud services.

\subsection{Ablation Study}

\begin{table}[t]
    \centering
    \caption{\textbf{Ablation Study.} Results are averaged across four horizons ($T \in \{64, 128, 192, 256\}$). We compare the full model (\textbf{SWIFT}) against: \textbf{w/o LC-Wavelet} (replaces learnable wavelet with standard convolution); \textbf{w/o MIM} (removes the entire module); and \textbf{w/o MIM-spatial/feature} (removes the respective stage). \best{Red} indicates the best result. Full details in Table~\ref{ablation-append}.}
    \label{tab:ablation}
    
    \renewcommand{\arraystretch}{1.1} 
    \setlength{\tabcolsep}{3pt}       

    \begin{adjustbox}{width=\linewidth}
    \begin{tabular}{c| cc | cc | cc | cc}
        \toprule
        \textbf{Datasets} & 
        \multicolumn{2}{c|}{\textbf{Ali2021}} & 
        \multicolumn{2}{c|}{\textbf{Ali2022}} & 
        \multicolumn{2}{c|}{\textbf{Fisher}} & 
        \multicolumn{2}{c}{\textbf{Google}} \\
        
        \cmidrule(r){1-1} \cmidrule(lr){2-3} \cmidrule(lr){4-5} \cmidrule(lr){6-7} \cmidrule(l){8-9}
        \textbf{Metric} & RMSE & MAE & RMSE & MAE & RMSE & MAE & RMSE & MAE \\
        \midrule
        \textbf{SWIFT} & \best{0.1084} & \best{0.0793} & \best{0.0479 } & \best{0.0321} & \best{0.0737} & \best{0.0495} & \best{0.0610 } & \best{0.0400} \\
        
        w/o LC-Wavelet      & 0.1400 & 0.0922 & 0.0501 & 0.0342 & 0.0771 & 0.0521 & 0.0639 & 0.0434 \\
        w/o MIM           & 0.1231 & 0.0855 & 0.0493 & 0.0334 & 0.0757 & 0.0509 & 0.0621 & 0.0416 \\
        w/o MIM-spatial   & 0.1188 & 0.0849 & 0.0491 & 0.0331 & 0.0754 & 0.0508 & 0.0617 & 0.0413 \\
        w/o MIM-feature   & 0.1159 & 0.0836 & 0.0489 & 0.0329 & 0.0751 & 0.0504 & 0.0616 & 0.0414 \\
        \bottomrule
    \end{tabular}
    \end{adjustbox}
\end{table}

\paragraph{Setup.}
To verify the effectiveness of each component, we construct three variants: 
(1) \textbf{w/o LC-Wavelet} replaces the learnable cascaded wavelet with a standard convolution ($k=3$) to validate multi-scale modeling; 
(2) \textbf{w/o MIM} removes the entire interaction module to test spatial dependencies;
(3) \textbf{w/o MIM-spatial/feature} removes only the inter- or intra-variable stages to decouple their effects.

\paragraph{Results.}
Table~\ref{tab:ablation} confirms the contribution of each module. \textbf{First}, \textbf{w/o LC-Wavelet} suffers the most severe degradation, identifying the LC-Wavelet as the core backbone. \textbf{Second}, \textbf{w/o MIM} consistently yields higher errors, underlining the necessity of modeling dependencies. \textbf{Finally}, \textbf{w/o MIM-spatial} generally underperforms \textbf{w/o MIM-feature}, suggesting that inter-variable Spatial Interaction is the dominant factor.

\subsection{Universality Verification}
\begin{table}[t!]
    \centering
    \caption{\textbf{Universality Verification on ETT Benchmarks.} Results are averaged across four horizons ($T \in \{96, 192, 336, 720\}$). \best{Red} indicates the best result. See Table~\ref{long-term-ett-append} in Appendix for full results.}
    \label{long-term-ett}
    
    \renewcommand{\arraystretch}{1.1} 
    \setlength{\tabcolsep}{3pt}       

    \begin{adjustbox}{width=\linewidth}
    \begin{tabular}{c|cc|cc|cc|cc}
        \toprule
        \textbf{Models} & \multicolumn{2}{c|}{ETTh1} & \multicolumn{2}{c|}{ETTh2} & \multicolumn{2}{c|}{ETTm1} & \multicolumn{2}{c}{ETTm2} \\
       \cmidrule(r){1-1} \cmidrule(lr){2-3} \cmidrule(lr){4-5} \cmidrule(lr){6-7} \cmidrule(l){8-9}
        \textbf{Metric} & MSE & MAE & MSE & MAE & MSE & MAE & MSE & MAE \\
        \midrule
        \textbf{SWIFT} & \best{0.429} & \best{0.428} & \best{0.381} & \best{0.401} & \best{0.729} & \best{0.576} & \best{1.016} & \best{0.700} \\
        TimesNet & 0.488 & 0.471 & 0.409 & 0.423 & 0.772 & 0.609 & 1.045 & 0.719 \\
        TimeXer & 0.444 & 0.435 & 0.382 & 0.402 & 0.759 & 0.596 & 1.048 & 0.717 \\
        Fremer & 0.442 & 0.435 & 0.387 & 0.406 & 0.761 & 0.605 & 1.068 & 0.724 \\
        WDFormer & 0.439 & 0.439 & 0.386 & 0.410 & 0.741 & 0.589 & 1.035 & 0.715 \\
        ModernTCN & 0.446 & 0.443 & 0.397 & 0.412 & 0.734 & 0.578 & 1.029 & 0.702 \\
        MICN & 0.474 & 0.460 & 0.404 & 0.433 & 0.774 & 0.620 & 1.073 & 0.734 \\
        \bottomrule
    \end{tabular}
    \end{adjustbox}
\end{table}

\paragraph{Setup.} 
We use the ETT benchmarks to verify the universality of SWIFT beyond cloud workload domains. To ensure a fair comparison with existing baselines, we strictly follow the standard experimental protocol~\cite{informer} by fixing the input length $L=96$ and evaluating across $T \in \{96, 192, 336, 720\}$.

\paragraph{Results.} 
The generalization results on the ETT benchmarks are summarized in Table~\ref{long-term-ett}. SWIFT achieves sota performance in \textbf{8 out of 8} cases and reduces MSE by up to \textbf{12.08\%} and MAE by up to \textbf{9.21\%}. These results verify SWIFT’s universality for general time-series forecasting beyond cloud workload domains.

\subsection{Sensitivity Analysis on Cascade Depth}

\begin{figure}[t]
    \centering
    \includegraphics[width=\linewidth]{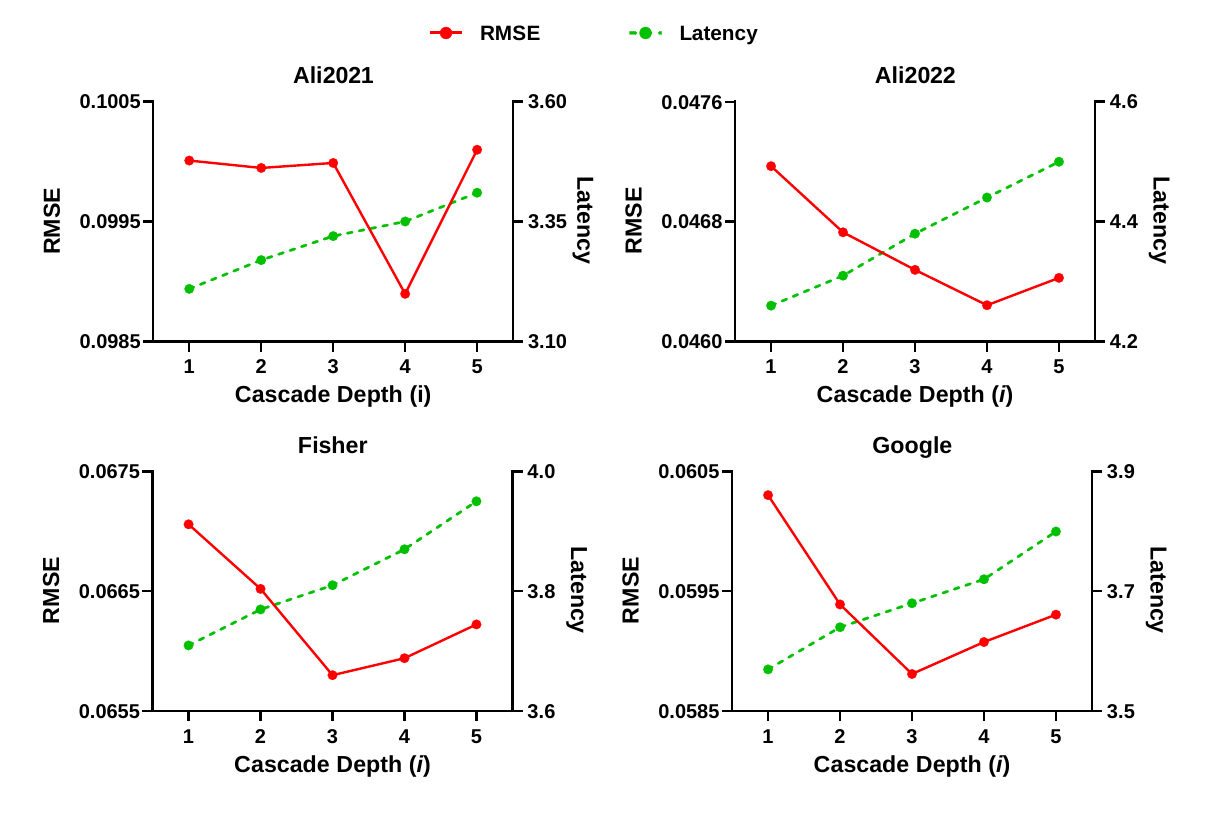}
    \caption{\textbf{Sensitivity Analysis of Cascade Depth ($i$).} Visualizing the trade-off between RMSE (red, left) and Latency (green, right) with fixed input length $L=512$ and prediction horizon $T=64$. While latency increases linearly, $i=4$ is optimal for fine-grained Ali datasets, whereas $i=3$ fits Fisher and Google.}
    \label{fig:sensitivity}
\end{figure}

\paragraph{Setup.}
To evaluate the trade-off between prediction accuracy (RMSE) and inference latency, we conduct a sensitivity analysis on the cascade depth $i$. We vary the depth $i \in \{1, 2, 3, 4, 5\}$ while fixing the input length at $L=512$ and the prediction horizon at $T=64$.

\paragraph{Results.}
Figure~\ref{fig:sensitivity} reveals two key observations. First, regarding efficiency, the inference latency increases linearly with the cascade depth. Second, regarding accuracy, the optimal depth proves to be domain-dependent due to varying sampling intervals. Specifically, the Ali datasets (recorded at high frequencies) require deeper decomposition ($i=4$) to capture fine-grained dynamics, whereas datasets with coarser intervals (Fisher and Google) peak at a shallower depth ($i=3$). Consequently, we adopt $i=4$ for the Ali datasets and $i=3$ for the others to achieve the optimal balance between performance and cost.

\section{Conclusion}
\label{section:5}
In this paper, we propose SWIFT, a Spatio-temporal Wavelet Integrated Forecasting Framework
for Workload Traces. By leveraging a Learnable Cascaded Wavelet Path and a Multivariate Interaction Module, SWIFT effectively captures non-stationary multi-scale periodicities and spatial correlations. Crucially, to ensure deployment efficiency, we employed a hardware-friendly pure convolutional architecture with linear $O(L)$ complexity. By replacing recursive operations with GPU-native convolutions, SWIFT achieves sota accuracy with superior inference throughput.
\section*{Impact Statement}
This work introduces a framework for cloud workload forecasting, where the achieved improvements in both accuracy and efficiency primarily enable proactive auto-scaling to enhance service stability and user experience in modern cloud environments. Beyond direct cloud applications, these improved forecasting capabilities extend to enhancing resource allocation efficiency in sectors such as edge computing, energy grids, and logistics, thereby reducing operational costs. However, improved accuracy does not guarantee stability; to mitigate risks associated with overreliance on automated scaling during unseen anomalies, it is crucial to implement fail-safe mechanisms and adhere to data privacy standards ensuring responsible deployment.

\bibliography{Reference}

@article{1-1,
	title={Kubernetes-oriented microservice placement with dynamic resource allocation},
	author={Ding, Zhijun and Wang, Song and Jiang, Changjun},
	journal={IEEE Transactions on Cloud Computing},
	year={2022},
	publisher={IEEE}
}

@INPROCEEDINGS{1-2,
	author={Dua, Rajdeep and Raja, A Reddy and Kakadia, Dharmesh},
	booktitle={2014 IEEE International Conference on Cloud Engineering}, 
	title={Virtualization vs Containerization to Support PaaS}, 
	year={2014},
	volume={},
	number={},
	pages={610-614},
	doi={10.1109/IC2E.2014.41}}

@inproceedings{1-3,
	title={Predictive Auto-scaling: LSTM-Based Multi-step Cloud Workload Prediction},
	author={Suleiman, Basem and Alibasa, Muhammad Johan and Chang, Ya-Yuan and Anaissi, Ali},
	booktitle={International Conference on Service-Oriented Computing},
	pages={5--16},
	year={2023},
	organization={Springer}
}

@article{1-4,
  title={Proactive auto-scaling for cloud environments using temporal convolutional neural networks},
  author={Golshani, Ehsan and Ashtiani, Mehrdad},
  journal={Journal of Parallel and Distributed Computing},
  volume={154},
  pages={119--141},
  year={2021},
  publisher={Elsevier}
}

@article{1-5,
  title={Proactive failure-aware task scheduling framework for cloud computing},
  author={Alahmad, Yanal and Daradkeh, Tariq and Agarwal, Anjali},
  journal={IEEE Access},
  volume={9},
  pages={106152--106168},
  year={2021},
  publisher={IEEE}
}

@article{1-6,
  title={A purely proactive scheduling procedure for the resource-constrained project scheduling problem with stochastic activity durations},
  author={Lamas, Patricio and Demeulemeester, Erik},
  journal={Journal of Scheduling},
  volume={19},
  number={4},
  pages={409--428},
  year={2016},
  publisher={Springer}
}

@book{mallat1999wavelet,
  title={A wavelet tour of signal processing},
  author={Mallat, Stephane},
  year={1999},
  publisher={Academic Press}
}

@inproceedings{finder2024wavelet,
  title={Wavelet convolutions for large receptive fields},
  author={Finder, Shahaf E and Amoyal, Roy and Treister, Eran and Freifeld, Oren},
  booktitle={European Conference on Computer Vision},
  pages={363--380},
  year={2024},
  organization={Springer}
}

@article{wu2022timesnet,
  title={Timesnet: Temporal 2d-variation modeling for general time series analysis},
  author={Wu, Haixu and Hu, Tengge and Liu, Yong and Zhou, Hang and Wang, Jianmin and Long, Mingsheng},
  journal={arXiv preprint arXiv:2210.02186},
  year={2022}
}

@article{frets,
  title={Frequency-domain MLPs are more effective learners in time series forecasting},
  author={Yi, Kun and Zhang, Qi and Fan, Wei and Wang, Shoujin and Wang, Pengyang and He, Hui and An, Ning and Lian, Defu and Cao, Longbing and Niu, Zhendong},
  journal={Advances in Neural Information Processing Systems},
  volume={36},
  pages={76656--76679},
  year={2023}
}

@inproceedings{fredformer,
  title={Fredformer: Frequency debiased transformer for time series forecasting},
  author={Piao, Xihao and Chen, Zheng and Murayama, Taichi and Matsubara, Yasuko and Sakurai, Yasushi},
  booktitle={Proceedings of the 30th ACM SIGKDD conference on knowledge discovery and data mining},
  pages={2400--2410},
  year={2024}
}

@inproceedings{wdformer,
  title={WDformer: A Wavelet-based Differential Transformer Model for Time Series Forecasting},
  author={Wang, Xiaojian and Zhang, Chaoli and Zheng, Zhonglong and Jiang, Yunliang},
  booktitle={Proceedings of the 34th ACM International Conference on Information and Knowledge Management},
  pages={3093--3102},
  year={2025}
}

@article{MODWT-LSTM,
  title={Novel wavelet-LSTM approach for time series prediction},
  author={Tamilselvi, C and Paul, Ranjit Kumar and Yeasin, Md and Paul, AK},
  journal={Neural Computing and Applications},
  volume={37},
  number={17},
  pages={10521--10530},
  year={2025},
  publisher={Springer}
}

@article{liao2020multiple,
  title={Multiple wavelet convolutional neural network for short-term load forecasting},
  author={Liao, Zhifang and Pan, Haihui and Fan, Xiaoping and Zhang, Yan and Kuang, Li},
  journal={IEEE Internet of Things Journal},
  volume={8},
  number={12},
  pages={9730--9739},
  year={2020},
  publisher={IEEE}
}

@inproceedings{informer,
  title={Informer: Beyond efficient transformer for long sequence time-series forecasting},
  author={Zhou, Haoyi and Zhang, Shanghang and Peng, Jieqi and Zhang, Shuai and Li, Jianxin and Xiong, Hui and Zhang, Wancai},
  booktitle={Proceedings of the AAAI conference on artificial intelligence},
  volume={35},
  number={12},
  pages={11106--11115},
  year={2021}
}

@article{timexer,
  title={Timexer: Empowering transformers for time series forecasting with exogenous variables},
  author={Wang, Yuxuan and Wu, Haixu and Dong, Jiaxiang and Qin, Guo and Zhang, Haoran and Liu, Yong and Qiu, Yunzhong and Wang, Jianmin and Long, Mingsheng},
  journal={Advances in Neural Information Processing Systems},
  volume={37},
  pages={469--498},
  year={2024}
}

@article{autoformer,
  title={Autoformer: Decomposition transformers with auto-correlation for long-term series forecasting},
  author={Wu, Haixu and Xu, Jiehui and Wang, Jianmin and Long, Mingsheng},
  journal={Advances in neural information processing systems},
  volume={34},
  pages={22419--22430},
  year={2021}
}

@inproceedings{zhou2022fedformer,
  title={Fedformer: Frequency enhanced decomposed transformer for long-term series forecasting},
  author={Zhou, Tian and Ma, Ziqing and Wen, Qingsong and Wang, Xue and Sun, Liang and Jin, Rong},
  booktitle={International conference on machine learning},
  pages={27268--27286},
  year={2022},
  organization={PMLR}
}

@inproceedings{crossformer,
  title={Crossformer: Transformer utilizing cross-dimension dependency for multivariate time series forecasting},
  author={Zhang, Yunhao and Yan, Junchi},
  booktitle={The eleventh international conference on learning representations},
  year={2023}
}

@article{itransformer,
  title={itransformer: Inverted transformers are effective for time series forecasting},
  author={Liu, Yong and Hu, Tengge and Zhang, Haoran and Wu, Haixu and Wang, Shiyu and Ma, Lintao and Long, Mingsheng},
  journal={arXiv preprint arXiv:2310.06625},
  year={2023}
}

@article{fremer,
  title={Fremer: Lightweight and Effective Frequency Transformer for Workload Forecasting in Cloud Services},
  author={Chen, Jiadong and Ye, Hengyu and Jiang, Fuxin and He, Xiao and Zhang, Tieying and Chen, Jianjun and Gao, Xiaofeng},
  journal={arXiv preprint arXiv:2507.12908},
  year={2025}
}

@inproceedings{moderntcn,
  title={Moderntcn: A modern pure convolution structure for general time series analysis},
  author={Luo, Donghao and Wang, Xue},
  booktitle={The twelfth international conference on learning representations},
  pages={1--43},
  year={2024}
}

@inproceedings{micn,
  title={Micn: Multi-scale local and global context modeling for long-term series forecasting},
  author={Wang, Huiqiang and Peng, Jian and Huang, Feihu and Wang, Jince and Chen, Junhui and Xiao, Yifei},
  booktitle={The eleventh international conference on learning representations},
  year={2023}
}

@inproceedings{group,
  title={Group: An end-to-end multi-step-ahead workload prediction approach focusing on workload group behavior},
  author={Feng, Binbin and Ding, Zhijun},
  booktitle={Proceedings of the ACM Web Conference 2023},
  pages={3098--3108},
  year={2023}
}

@inproceedings{5-1,
	title={Characterizing microservice dependency and performance: Alibaba trace analysis},
	author={Luo, Shutian and Xu, Huanle and Lu, Chengzhi and Ye, Kejiang and Xu, Guoyao and Zhang, Liping and Ding, Yu and He, Jian and Xu, Chengzhong},
	booktitle={Proceedings of the ACM Symposium on Cloud Computing},
	pages={412--426},
	year={2021}
}

@inproceedings{5-2,
	title={The power of prediction: microservice auto scaling via workload learning},
	author={Luo, Shutian and Xu, Huanle and Ye, Kejiang and Xu, Guoyao and Zhang, Liping and Yang, Guodong and Xu, Chengzhong},
	booktitle={Proceedings of the 13th Symposium on Cloud Computing},
	pages={355--369},
	year={2022}
}

@inproceedings{5-3,
	title={Fisher: An efficient container load prediction model with deep neural network in clouds},
	author={Tang, Xuehai and Liu, Qiuyang and Dong, Yangchen and Han, Jizhong and Zhang, Zhiyuan},
	booktitle={2018 IEEE Intl Conf on Parallel \& Distributed Processing with Applications, Ubiquitous Computing \& Communications, Big Data \& Cloud Computing, Social Computing \& Networking, Sustainable Computing \& Communications (ISPA/IUCC/BDCloud/SocialCom/SustainCom)},
	pages={199--206},
	year={2018},
	organization={IEEE}
}

@inproceedings{5-4,
	author = {Verma, Abhishek and Pedrosa, Luis and Korupolu, Madhukar and Oppenheimer, David and Tune, Eric and Wilkes, John},
	title = {Large-scale cluster management at Google with Borg},
	year = {2015},
	isbn = {9781450332385},
	publisher = {Association for Computing Machinery},
	address = {New York, NY, USA},
	url = {https://doi.org/10.1145/2741948.2741964},
	doi = {10.1145/2741948.2741964},
	booktitle = {Proceedings of the Tenth European Conference on Computer Systems},
	articleno = {18},
	numpages = {17},
	location = {Bordeaux, France},
	series = {EuroSys '15}
}

@article{paszke2019pytorch,
  title={Pytorch: An imperative style, high-performance deep learning library},
  author={Paszke, Adam and Gross, Sam and Massa, Francisco and Lerer, Adam and Bradbury, James and Chanan, Gregory and Killeen, Trevor and Lin, Zeming and Gimelshein, Natalia and Antiga, Luca and others},
  journal={Advances in neural information processing systems},
  volume={32},
  year={2019}
}

@article{ding2022coin,
  title={COIN: A container workload prediction model focusing on common and individual changes in workloads},
  author={Ding, Zhijun and Feng, Binbin and Jiang, Changjun},
  journal={IEEE Transactions on Parallel and Distributed Systems},
  volume={33},
  number={12},
  pages={4738--4751},
  year={2022},
  publisher={IEEE}
}

@article{zhao2024tfegru,
  title={TFEGRU: time-frequency enhanced gated recurrent unit with attention for cloud workload prediction},
  author={Zhao, Feiyu and Lin, Weiwei and Lin, Shengsheng and Zhong, Haocheng and Li, Keqin},
  journal={IEEE Transactions on Services Computing},
  year={2024},
  publisher={IEEE}
}
\bibliographystyle{icml2026}

\newpage
\appendix
\onecolumn
\section{Experimental Details}

\subsection{Datasets}
\label{sec: append-datasets}
\begin{table}[t!]
    \centering
    \caption{\textbf{Dataset Statistics.} We evaluate our model on eight datasets covering various domains.}
    \label{tab:dataset_stat}
    
    \renewcommand{\arraystretch}{1} 
    \setlength{\tabcolsep}{5pt}        

    \begin{adjustbox}{width=\linewidth}
    \begin{tabular}{c|ccccccccc}
    \toprule
    Dataset & Ali2021 & Ali2022 & Fisher & Google & ETTh1 & ETTh2 & ETTm1 & ETTm2 \\
    \midrule
    Dataset Size & 1440 & 18720 & 78000 & 8352 & 17420 & 17420 & 69680 & 69680 \\
    Variable Number & 20000 & 28000 & 10 & 15000 & 7 & 7 & 7 & 7 \\
    Sampling Frequency & 30s & 1 min & - & 5 mins & 1 hour & 1 hour & 15 mins & 15 mins \\
    Domain & Cloud Workload & Cloud Workload & Cloud Workload & Cloud Workload & Electricity & Electricity & Electricity & Electricity \\
    \bottomrule
  \end{tabular}
    \end{adjustbox}
\end{table}

We evaluate the performance of our proposed method and all baselines on four large-scale real-world workload datasets: Ali2021, Ali2022, Fisher, and Google. Our experiments cover comprehensive aspects including long-term and short-term workload forecasting accuracy, generalization ability, and computational efficiency. Furthermore, to verify the universality of our model beyond the specific domain of workload prediction, we also conduct experiments on the widely used ETT datasets (ETTh1, ETTh2, ETTm1, ETTm2). The detailed statistics of all datasets, including dataset size (total timesteps), variable number, and sampling frequency, are summarized in Table~\ref{tab:dataset_stat}.

We split all datasets into training, validation, and test sets in chronological order by the ratio of 6:2:2 for the ETT dataset and 7:1:2 for the other datasets. The training, validation, and test sets are zero-mean normalized with the mean and standard deviation of the training set. Each of the above datasets only contains one continuous long time series, and we obtain samples by sliding window.

Detailed descriptions of the datasets are as follows:

\begin{description}
    \item[Ali2021 (Alibaba cluster-trace-microservices-v2021)] provides usage data of real cloud computing systems, including more than 20,000 microservices over 12 hours in 2021.
    
    \item[Ali2022 (Alibaba cluster-trace-microservices-v2022)] provides usage data of real cloud computing systems, including more than 28,000 microservices during 13 days in 2022.

    \item[Fisher] provides usage data of a real online Kubernetes system, including 10 containers over 30 days.

    \item[Google (Google 2011 cluster data)] provides usage data of Google data center, including about 15,000 machines during 29 days in 2011.
    
    \item[ETT (Electricity Transformer Temperature)] contains the data collected from electricity transformers with 7 sensors, including load, oil temperature, etc. It contains two sub-datasets labeled with 1 and 2, corresponding to two different electric transformers from two separated counties in China. Each of them contains 2 different resolutions (15 minutes and 1 hour) denoted with m and h. Thus, in total we have 4 ETT datasets: ETTh1, ETTh2, ETTm1, and ETTm2.
\end{description}

\subsection{Implementation Details}
\label{sec: append-metric}
\paragraph{Implementation.} All experiments are repeated three times with different seeds, implemented in PyTorch~\cite{paszke2019pytorch}, and conducted on a single NVIDIA Tesla V100 16GB GPU. Our method is trained with the $L_2$ loss using the ADAM optimizer with an initial learning rate of $10^{-4}$. The batch size is set to 40. The training process is early stopped within three epochs if there is no loss degradation on the validation set. To initialize the predicted future values, we employ an MLP along the temporal dimension.

For a fair comparison with baselines, we fix the input length to 512 for all workload datasets and 96 for ETT datsets. All the reproduced baselines are implemented based on the configurations of the original papers or official codes. We keep the input embedding and the final projection layer consistent among different base models to solely evaluate the capability of the base models themselves.

\paragraph{Metrics.}
We employ Root Mean Square Error (RMSE) and Mean Absolute Error (MAE) for general workload prediction tasks. Specifically for short-term workload forecasting, we additionally utilize Symmetric Mean Absolute Percentage Error (SMAPE) alongside RMSE and MAE to capture relative prediction accuracy in low-load regions. Furthermore, to verify the universality of our model beyond the specific domain of workload prediction, we follow the standard protocol~\cite{informer} on ETT datasets and use Mean Squared Error (MSE) and Mean Absolute Error (MAE).

\begin{equation}
    \text{MSE} = \frac{1}{N} \sum_{i=1}^{N} (y_i - \hat{y}_i)^2
    \label{eq:mse}
\end{equation}

\begin{equation}
    \text{RMSE} = \sqrt{\frac{1}{N} \sum_{i=1}^{N} (y_i - \hat{y}_i)^2}
    \label{eq:rmse}
\end{equation}

\begin{equation}
    \text{MAE} = \frac{1}{N} \sum_{i=1}^{N} |y_i - \hat{y}_i|
    \label{eq:mae}
\end{equation}

\begin{equation}
    \text{SMAPE} = \frac{100\%}{N} \sum_{i=1}^{N} \frac{|y_i - \hat{y}_i|}{(|y_i| + |\hat{y}_i|) / 2}
    \label{eq:smape}
\end{equation}

\section{Complete Results and Analysis}

\subsection{Long-Term Workload Forecasting}
\begin{table*}[h]
    \centering
    \caption{\textbf{Long-Term Workload Forecasting Results}. We report the full performance breakdown across all four prediction horizons $T \in \{64, 128, 192, 256\}$. Avg is averaged from all four prediction lengths. \best{Red} indicates the best result.}
    \label{long-term-append}
    
    \renewcommand{\arraystretch}{1.15} 
    \setlength{\tabcolsep}{3pt}        

    \begin{adjustbox}{width=\linewidth}
    \begin{tabular}{cc|cc|cc|cc|cc|cc|cc|cc}
        \toprule
        \multicolumn{2}{c|}{Models} & 
        \multicolumn{2}{c|}{\textbf{SWIFT}} & 
        \multicolumn{2}{c}{TimesNet} &   
        \multicolumn{2}{c}{TimeXer} &       
        \multicolumn{2}{c}{Fremer} &       
        \multicolumn{2}{c}{WDFormer} &        
        \multicolumn{2}{c}{ModernTCN} &    
        \multicolumn{2}{c}{MICN} \\    
        
        \cmidrule(r){1-2} \cmidrule(lr){3-4} \cmidrule(lr){5-6} \cmidrule(lr){7-8} \cmidrule(lr){9-10} \cmidrule(lr){11-12} \cmidrule(lr){13-14} \cmidrule(l){15-16}
        \multicolumn{2}{c|}{Metric} & RMSE & MAE & RMSE & MAE & RMSE & MAE & RMSE & MAE & RMSE & MAE & RMSE & MAE & RMSE & MAE \\
        \midrule

        \multirow{4}{*}{\rotatebox{90}{Ali2021}}
        & 64  & \best{0.0988} & \best{0.0694} & 0.1152 & 0.0851 & 0.1005 & 0.0713 & 0.1009 & 0.0719 & 0.1003 & 0.0709 & 0.1018 & 0.0724 & 0.0991 & 0.0706 \\
        & 128 & \best{0.1042} & \best{0.0766} & 0.1218 & 0.0911 & 0.1076 & 0.0785 & 0.1072 & 0.0782 & 0.1081 & 0.0786 & 0.1087 & 0.0790 & 0.1051 & 0.0774 \\
        & 192 & 0.1122 & 0.0835 & 0.1243 & 0.0924 & 0.1143 & 0.0848 & 0.1144 & 0.0847 & 0.1138 & 0.0846 & 0.1153 & 0.0852 & \best{0.1105} & \best{0.0834} \\
        & 256 & 0.1183 & 0.0875 & 0.1239 & 0.0918 & 0.1188 & 0.0775 & 0.1189 & \best{0.0874} & 0.1186 & 0.0879 & 0.1215 & 0.0890 & \best{0.1175} & 0.0887 \\
        & Avg & 0.1084 & \best{0.0793} & 0.1213 & 0.0901 & 0.1103 & 0.0805 & 0.1104 & 0.0805 & 0.1101 & 0.0805 & 0.1118 & 0.0814 & \best{0.1080} & 0.0800\\
        \midrule

        \multirow{4}{*}{\rotatebox{90}{Ali2022}} 
        & 64  & \best{0.0462} & \best{0.0307} & 0.0851 & 0.0344 &  0.0485 &  0.0350 &  0.0464 &  0.0308 &  0.0470 &  0.0312 &  0.0466 &  0.0310 &  0.0470 &  0.0311 \\
        & 128 & \best{0.0475} & \best{0.0318} & 0.0911 &  0.0345 &  0.0497 &  0.0359 &  0.0479 &  0.0321 &  0.0484 &  0.0325 &  0.0479 &  0.0321 &  0.0487 &  0.0326 \\
        & 192 & \best{0.0486} & \best{0.0326} & 0.0924 &  0.0349 &  0.0506 &  0.0368 &  0.0489 &  0.0330 &  0.0494 &  0.0334 &  0.0489 &  0.0329 &  0.0497 &  0.0339 \\
        & 256 & \best{0.0492} & \best{0.0333} & 0.0918 &  0.0351 &  0.0516 &  0.0376 &  0.0497 &  0.0337 &  0.0501 &  0.0341 &  0.0498 &  0.0336 &  0.0505 &  0.0345 \\
        & Avg & \best{0.0479} & \best{0.0321} & 0.0901 &  0.0347 &  0.0501 &  0.0363 &  0.0482 &  0.0324 &  0.0487 &  0.0328 &  0.0483 &  0.0324 &  0.0490 &  0.0330 \\
        \midrule

        \multirow{4}{*}{\rotatebox{90}{Fisher}} 
        & 64  & 0.0658 & \best{0.0412} & 0.0704 &  0.0484 &  \best{0.0656} &  0.04216 &  0.0691 &  0.0443 &  0.0722 &  0.0484 &  0.0669 &  0.0669 &  0.0681 &  0.0426 \\
        & 128 & \best{0.0743} & 0.0503 & 0.0766 &  0.0529 &  0.0769 &  0.05116 &  0.0765 & \best{0.0499} &  0.0763 &  0.0511 &  0.0764 &  0.0764 &  0.0756 &  0.0500 \\
        & 192 & \best{0.0763} & \best{0.0524} & 0.0804 &  0.0569 &  0.0780 &  0.05255 &  0.0778 &  0.0526 &  0.0815 &  0.0563 &  0.0772 &  0.0772 &  0.0773 &  0.0534 \\
        & 256 & \best{0.0783} & \best{0.0541} & 0.0813 &  0.0571 &  0.0803 &  0.05537 &  0.0793 &  0.0543 &  0.0825 &  0.0582 &  0.0791 &  0.0791 &  0.0806 &  0.0554 \\
        & Avg & \best{0.0737} & \best{0.0495} &  0.0772 &  0.0538 &  0.0752 &  0.05031 &  0.0757 &  0.0503 &  0.0781 &  0.0535 &  0.0749 &  0.0749 &  0.0754 &  0.0503 \\
        \midrule

        \multirow{4}{*}{\rotatebox{90}{Google}} 
        & 64  & \best{0.0588} & \best{0.0373} &  0.0629 &  0.0391 &  0.0593 &  0.0391 &  0.0593 &  0.0388 &  0.0613 &  0.0408 &  0.0591 &  0.03858 &  0.0635 &  0.0436 \\
        & 128 & 0.0609 & \best{0.0396} & 0.0647 &  0.0407 &  0.0613 &  0.0407 &  \best{0.0609} &  0.0399 &  0.0631 &  0.0422 &  0.0613 &  0.0401 &  0.0668 &  0.0459 \\
        & 192 & \best{0.0616} & 0.0411 &  0.0661 &  0.0419 &  0.0628 &  0.0419 &  0.0618 &  \best{0.0409} &  0.0645 &  0.0434 &  0.0626 &  0.0411 &  0.0671 &  0.0463 \\
        & 256 & \best{0.0625} & 0.0422 & 0.0675 &  0.0426 &  0.0638 &  0.0426 &  0.0627 &  \best{0.0416} &  0.0655 &  0.0442 &  0.0641 &  0.0423 &  0.0685 &  0.0484 \\
        & Avg & \best{0.0610} & \best{0.0400} & 0.0653 &  0.0411 &  0.0618 &  0.0411 &  0.0612 &  0.0403 &  0.0636 &  0.0427 &  0.0618 &  0.0405 &  0.0665 &  0.0461 \\
        \midrule

        \multicolumn{2}{c|}{1st Count} & \best{15} & \best{15} & 0 & 0 & 1 & 0 & 1 & 4 & 0 & 0 & 0 & 0 & 3 & 1 \\
        \bottomrule
    \end{tabular}
    \end{adjustbox}
\end{table*}

\paragraph{Detailed Long-Term Workload Forecasting Results.} The full results of long-term workload forecasting are presented in Table~\ref{long-term-append}. Overall, our proposed SWIFT demonstrates superior performance across four large-scale datasets, achieving the best results in \textbf{30 out of 40} cases (covering both RMSE and MAE metrics across four prediction lengths).

\subsection{Intra-Dataset Generalization}
\begin{table*}[h]
    \centering
    \caption{\textbf{Detailed Intra-Dataset Generalization Results.} Models are trained on the first 50\% disjoint subset of each dataset and evaluated on the remaining 50\% unseen subset. We report the full performance breakdown across all four prediction horizons $T \in \{64, 128, 192, 256\}$. Avg is averaged from all four prediction lengths. \best{Red} indicates the best result.}
    \label{intra-data-append}
    
    \renewcommand{\arraystretch}{1.15} 
    \setlength{\tabcolsep}{3pt}        

    \begin{adjustbox}{width=\linewidth}
    \begin{tabular}{cc|cc|cc|cc|cc|cc|cc|cc}
        \toprule
        \multicolumn{2}{c|}{Models} & 
        \multicolumn{2}{c|}{\textbf{SWIFT}} & 
        \multicolumn{2}{c}{TimesNet} &   
        \multicolumn{2}{c}{TimeXer} &       
        \multicolumn{2}{c}{Fremer} &       
        \multicolumn{2}{c}{WDFormer} &        
        \multicolumn{2}{c}{ModernTCN} &    
        \multicolumn{2}{c}{MICN} \\    
        
        \cmidrule(r){1-2} \cmidrule(lr){3-4} \cmidrule(lr){5-6} \cmidrule(lr){7-8} \cmidrule(lr){9-10} \cmidrule(lr){11-12} \cmidrule(lr){13-14} \cmidrule(l){15-16}
        \multicolumn{2}{c|}{Metric} & RMSE & MAE & RMSE & MAE & RMSE & MAE & RMSE & MAE & RMSE & MAE & RMSE & MAE & RMSE & MAE \\
        \midrule

        \multirow{4}{*}{\rotatebox{90}{Ali2021}}
        & 64  & 0.0693 & \best{0.0513} & 0.0736 & 0.0566 &  0.0703 &  0.0539 &  0.0699 &  0.0522 &  0.0761 &  0.0589 &  \best{0.0688} &  0.0519 &  0.0743 &  0.0567 \\
        & 128 & \best{0.0695} & \best{0.0519} & 0.0753 &  0.0577 &  0.0727 &  0.0557 &  0.0719 &  0.0542 &  0.0768 &  0.0592 &  0.0709 &  0.0530 &  0.0762 &  0.0563 \\
        & 192 & \best{0.0692} & \best{0.0525} & 0.0745 &  0.0574 &  0.0732 &  0.0567 &  0.0726 &  0.0554 &  0.0765 &  0.0593 &  0.0702 &  0.0531 &  0.0748 &  0.0583 \\
        & 256 & 0.0705 & 0.0537 & 0.0771 &  0.0586 &  0.0744 &  0.0580 &  0.0739 &  0.0569 &  0.0785 &  0.0610 &  \best{0.0699} &  \best{0.0532} &  0.0729 &  0.0598 \\
        & Avg & \best{0.0696} & \best{0.0523} & 0.0751 &  0.0576 &  0.0726 &  0.0561 &  0.0721 &  0.0547 &  0.0769 &  0.0596 &  0.0700 &  0.0528 &  0.0746 &  0.0578\\
        \midrule

        \multirow{4}{*}{\rotatebox{90}{Ali2022}} 
        & 64  & 0.0395 &  0.0283 &  0.0401 &  0.0290 &  0.0395 &  \best{0.0281} &  \best{0.0394} &  0.0281 &  0.0399 &  0.0286 &  0.0399 &  0.0289 &  0.0405 &  0.0295 \\
        & 128 & 0.0401 &  \best{0.0285} &  0.0406 &  0.0296 &  0.0402 &  0.0289 &  \best{0.0399} &  0.0290 &  0.0404 &  0.0292 &  0.0403 &  0.0294 &  0.0413 &  0.0299 \\
        & 192 & \best{0.0404} & \best{0.0290} & 0.0410 &  0.0299 &  0.0407 &  0.0295 &  0.0406 &  0.0295 &  0.0409 &  0.0301 &  0.0407 &  0.0298 &  0.0420 &  0.0301 \\
        & 256 & \best{0.0408} & \best{0.0294} & 0.0415 &  0.0305 &  0.0411 &  0.0300 &  0.0410 &  0.0299 &  0.0413 &  0.0304 &  0.0411 &  0.0302 &  0.0421 &  0.0311 \\
        & Avg & \best{0.0402} & \best{0.0288} & 0.0408 &  0.0298 &  0.0404 &  0.0292 &  0.0402 &  0.0291 &  0.0406 &  0.0296 &  0.0405 &  0.0296 &  0.0415 &  0.0302 \\
        \midrule

        \multirow{4}{*}{\rotatebox{90}{Fisher}} 
        & 64  & \best{0.0539} &  \best{0.0384} &  0.0579 &  0.0417 &  0.0565 &  0.0403 &  0.0545 &  0.0389 &  0.0581 &  0.0426 &  0.0547 &  0.0391 &  0.0579 &  0.0410 \\
        & 128 &\best{0.0592} &  \best{0.0416} &  0.0615 &  0.0444 &  0.0615 &  0.0437 &  0.0599 &  0.0424 &  0.0649 &  0.0467 &  0.0597 &  0.0424 &  0.0616 &  0.0457 \\
        & 192 & \best{0.0641} & \best{0.0446} &  0.0755 &  0.0527 &  0.0667 &  0.0469 &  0.0652 &  0.0458 &  0.0702 &  0.0499 &  0.0644 &  0.0452 &  0.0749 &  0.0508 \\
        & 256 & \best{0.0684} & \best{0.0475} & 0.0719 &  0.0502 &  0.0713 &  0.0496 &  0.0694 &  0.0481 &  0.0748 &  0.0530 &  0.0684 &  0.0476 &  0.0757 &  0.0514 \\
        & Avg & \best{0.0614} & \best{0.0431} &   0.0667 &  0.0473 &  0.0640 &  0.0451 &  0.0623 &  0.0438 &  0.0670 &  0.0480 &  0.0619 &  0.0436 &  0.0675 &  0.0472  \\
        \midrule

        \multirow{4}{*}{\rotatebox{90}{Google}} 
        & 64  & 0.0407 &  0.0251 &  0.0417 &  0.0263 &  0.0405 &  \best{0.0249} &  \best{0.0405} &  0.0251 &  0.0424 &  0.0266 &  0.0409 &  0.0255 &  0.0420 &  0.0274 \\
        & 128 & 0.0420 &  0.0260 &  0.0428 &  0.0272 &  0.0422 &  \best{0.0259} &  \best{0.0418} &  0.0261 &  0.0441 &  0.0280 &  0.0422 &  0.0264 &  0.0432 &  0.0278 \\
        & 192 & \best{0.0421} &  \best{0.0265} &  0.0442 &  0.0285 &  0.0433 &  0.0268 &  0.0428 &  0.0269 &  0.0451 &  0.0287 &  0.0429 &  0.0269 &  0.0458 &  0.0292  \\
        & 256 & \best{0.0432} &  \best{0.0269} &  0.0448 &  0.0288 &  0.0439 &  0.0273 &  0.0435 &  0.0276 &  0.0455 &  0.0292 &  0.0436 &  0.0272 &  0.0459 &  0.0298 \\
        & Avg & \best{0.0420} & \best{0.0261} & 0.0434 &  0.0277 &  0.0425 &  0.0263 &  0.04214 &  0.0264 &  0.0443 &  0.0281 &  0.0424 &  0.0265 &  0.0443 &  0.0286 \\
        \midrule

        \multicolumn{2}{c|}{1st Count} & \best{14} & \best{16} & 0 & 0 & 0 & 3 & 4 & 0 & 0 & 0 & 2 & 1 & 0 & 0 \\
        \bottomrule
    \end{tabular}
    \end{adjustbox}
\end{table*}

\paragraph{Detailed Intra-Dataset Generalization Results.} 
The complete breakdown of intra-dataset generalization performance is presented in Table~\ref{intra-data-append}. In this rigorous evaluation setting, where models are trained on the first half and tested on the disjoint latter half of each dataset, SWIFT demonstrates exceptional robustness, securing the best results in \textbf{30 out of 40} cases, verifying its capability to handle distribution shifts over extended periods.

\subsection{Cross-Dataset Generalization}
\paragraph{Detailed Cross-Dataset Generalization Results.} 
Table~\ref{cross-data-append} details the performance in the cross-dataset evaluation scenario. In this challenging zero-shot setting, where the model is trained exclusively on \textbf{Ali2022} and evaluated directly on unseen datasets (Ali2021, Fisher, and Google) without any fine-tuning, SWIFT demonstrates remarkable transferability. It secures the best performance in \textbf{22 out of 30} cases, validating that our model captures universal temporal patterns rather than merely overfitting to the source domain's specific statistics.

\begin{table*}[h]
    \centering
    \caption{\textbf{Detailed Cross-Dataset Generalization Result.} Models are trained solely on \textbf{Ali2022} and directly evaluated on \textbf{Ali2021}, \textbf{Google}, and \textbf{Fisher} without fine-tuning. We report the full performance breakdown across all four prediction horizons $T \in \{64, 128, 192, 256\}$. Avg is averaged from all four prediction lengths. \best{Red} indicates the best result.}
    \label{cross-data-append}
    
    \renewcommand{\arraystretch}{1.15} 
    \setlength{\tabcolsep}{3pt}        

    \begin{adjustbox}{width=\linewidth}
    \begin{tabular}{cc|cc|cc|cc|cc|cc|cc|cc}
        \toprule
        \multicolumn{2}{c|}{Models} & 
        \multicolumn{2}{c|}{\textbf{SWIFT}} & 
        \multicolumn{2}{c}{TimesNet} &   
        \multicolumn{2}{c}{TimeXer} &       
        \multicolumn{2}{c}{Fremer} &       
        \multicolumn{2}{c}{WDFormer} &        
        \multicolumn{2}{c}{ModernTCN} &    
        \multicolumn{2}{c}{MICN} \\    
        
        \cmidrule(r){1-2} \cmidrule(lr){3-4} \cmidrule(lr){5-6} \cmidrule(lr){7-8} \cmidrule(lr){9-10} \cmidrule(lr){11-12} \cmidrule(lr){13-14} \cmidrule(l){15-16}
        \multicolumn{2}{c|}{Metric} & RMSE & MAE & RMSE & MAE & RMSE & MAE & RMSE & MAE & RMSE & MAE & RMSE & MAE & RMSE & MAE \\
        \midrule

        \multirow{4}{*}{\rotatebox{90}{Ali2021}}
        & 64  & \best{0.0737} &  \best{0.0536} &  0.0757 &  0.0555 &  0.0748 & 0.0538 &  0.0741 &  0.0539 &  0.0760 &  0.0554 &  0.0751 &  0.0551 &  0.0758 &  0.0550 \\
        & 128 & \best{0.0739} &  \best{0.0545} &  0.0752 &  0.0561 &  0.0758 &  0.0562 &  0.0740 &  0.0548 &  0.0753 &  0.0559 &  0.0754 &  0.0561 &  0.0760 &  0.0571 \\
        & 192 & 0.0740 &  0.0557 & 0.0744 &  0.0558 &  0.0748 &  0.0563 &  \best{0.0737} &  \best{0.0551} &  0.0741 &  0.0557 &  0.0757 &  0.0569 &  0.0757 &  0.0573 \\
        & 256 & \best{0.0749} &  \best{0.0563} &  0.0763 &  0.0580 &  0.0770 &  0.0582 &  0.0757 &  0.0572 &  0.0752 &  0.0569 &  0.0771 &  0.0586 &  0.0751 &  0.0582 \\
        & Avg & \best{0.0741} & \best{0.0550} & 0.0754 &  0.0563 &  0.0756 &  0.0561 &  0.0744 &  0.0552 &  0.0752 &  0.0560 &  0.0758 &  0.0567 &  0.0757 &  0.0569 \\
        \midrule

        \multirow{4}{*}{\rotatebox{90}{Fisher}} 
        & 64  & 0.0485 &  0.0338 &  0.0557 &  0.0387 &  \best{0.0484} &  \best{0.0334} &  0.0487 &  0.0337 &  0.0515 &  0.0363 &  0.0510 &  0.0355 &  0.0569 &  0.0382 \\
        & 128 &\best{0.0537} &  \best{0.0371} & 0.0608 &  0.0420 &  0.0546 &  0.0377 &  0.0542 &  0.0375 &  0.0565 &  0.0396 &  0.0559 &  0.0387 &  0.0583 &  0.0421 \\
        & 192 & \best{0.0580} & \best{0.0400} &  0.0652 &  0.0444 &  0.0594 &  0.0410 &  0.0587 &  0.0405 &  0.0604 &  0.0421 &  0.0597 &  0.0412 &  0.0623 &  0.0455 \\
        & 256 & \best{0.0618} & \best{0.0413} & 0.0689 &  0.0465 &  0.0631 &  0.0433 &  0.0623 &  0.0428 &  0.0638 &  0.0441 &  0.0631 &  0.0432 &  0.0692 &  0.0472 \\
        & Avg & \best{0.0555} & \best{0.0380} &   0.0627 &  0.0429 &  0.0564 &  0.0388 &  0.0560 &  0.0386 &  0.0581 &  0.0405 &  0.0574 &  0.0397 &  0.0617 &  0.0433 \\ 
        \midrule

        \multirow{4}{*}{\rotatebox{90}{Google}} 
        & 64  & \best{0.0583} &  \best{0.0400} &  0.0624 &  0.0439 &  0.0592 &  0.0409 &  0.0599 &  0.0412 &  0.0611 &  0.0425 &  0.0614 &  0.0424 &  0.0627 &  0.0460 \\
        & 128 & \best{0.0600} &  0.0416 &  0.0636 &  0.0449 &  0.0609 &  0.0424 &  0.0614 &  \best{0.0413} &  0.0625 &  0.0437 &  0.0625 &  0.0435 &  0.0664 &  0.0496 \\
        & 192 & \best{0.0612} &  0.0423 &  0.0645 &  0.0457 &  0.0619 &  0.0434 &  0.0622 &  \best{0.0422} &  0.0634 &  0.0445 &  0.0634 &  0.0443 &  0.0696 &  0.0510 \\
        & 256 & 0.0629 &  0.0426 &  0.0653 &  0.0462 &  \best{0.0624} &  0.0443 &  0.0628 &  \best{0.0426} &  0.0642 &  0.0452 &  0.0642 &  0.0451 &  0.0717 &  0.0533 \\
        & Avg & \best{0.0606} & \best{0.0416} & 0.0640 &  0.0452 &  0.0611 &  0.0428 &  0.0616 &  0.0418 &  0.0628 &  0.0440 &  0.0629 &  0.0438 &  0.0676 &  0.0500 \\ 
        \midrule

        \multicolumn{2}{c|}{1st Count} & \best{12} & \best{10} & 0 & 0 & 2 & 1 & 1 & 4 & 0 & 0 & 0 & 0 & 0 & 0 \\
        \bottomrule
    \end{tabular}
    \end{adjustbox}
\end{table*}

\begin{table}[H]
    \centering
    \caption{\textbf{Sensitivity Analysis of Block Numbers.} We fix the input length $L=512$ and prediction horizon $T=64$.}
    \label{tab:block-sensitivity}
    
    \renewcommand{\arraystretch}{1} 
    \setlength{\tabcolsep}{4pt}       

    \begin{adjustbox}{width=0.7\linewidth}
    \begin{tabular}{c|cc|cc|cc|cc}
        \toprule
        \multirow{2}{*}{\textbf{Dataset}} & 
        \multicolumn{2}{c|}{\textbf{Block = 1}} & 
        \multicolumn{2}{c|}{\textbf{Block = 2}} & 
        \multicolumn{2}{c|}{\textbf{Block = 3}} & 
        \multicolumn{2}{c}{\textbf{Block = 4}} \\
        
        \cmidrule(lr){2-3} \cmidrule(lr){4-5} \cmidrule(lr){6-7} \cmidrule(l){8-9}
        
        & RMSE & Latency & RMSE & Latency & RMSE & Latency & RMSE & Latency \\
        \midrule

        Ali2021 & 0.0989 & 3.35 & 0.1004 & 7.73 & 0.1006 & 12.77 & 0.1008 & 16.97 \\
        Ali2022 & 0.0462 & 4.47 & 0.0457 & 8.70 & 0.0462 & 13.28 & 0.0463 & 17.89 \\
        Fisher  & 0.0658 & 3.81 & 0.0658 & 6.45 & 0.0660 & 11.15 & 0.0663 & 16.35 \\
        Google  & 0.0588 & 3.68 & 0.0585 & 7.83 & 0.0589 & 13.84 & 0.0590 & 17.65 \\
        
        \bottomrule
    \end{tabular}
    \end{adjustbox}
\end{table}

\subsection{Sensitivity Analysis}
\paragraph{Impact of Block Numbers.}
To investigate the trade-off between forecasting accuracy and computational efficiency, we conduct a sensitivity analysis on the number of stacked blocks in our SWIFT architecture. We vary the model depth from 1 to 4 blocks while fixing the input sequence length at $L=512$ and the prediction horizon at $T=64$ for a consistent comparison. The results, summarized in Table~\ref{tab:block-sensitivity}, reveal two key observations. First, regarding \textbf{accuracy}, increasing the number of blocks yields negligible performance gains; in fact, the single-block model (Block = 1) achieves the lowest RMSE on both Ali2021 and Fisher datasets, while deeper models (Block = 3 or 4) tend to suffer from slight performance degradation due to potential overfitting. Second, regarding \textbf{efficiency}, the inference latency increases linearly with the number of blocks (e.g., jumping from 3.35 ms to 16.97 ms on Ali2021). Considering that Block = 1 offers competitive accuracy with the lowest computational cost, we adopt \textbf{Block = 1} as the default setting to ensure our model remains lightweight and efficient.

\begin{figure}[H]
    \centering
    \includegraphics[width=0.6\linewidth]{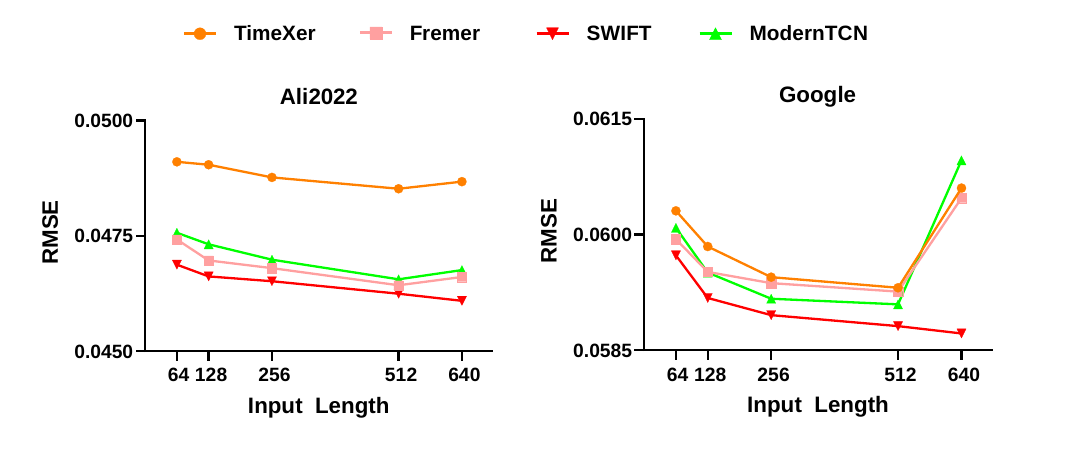}
    \caption{\textbf{Sensitivity Analysis of Input Length ($L$).} The RMSE results with different input lengths $L \in \{64, 128, 256, 512, 640\}$ with a fixed prediction horizon $T=64$.}
    \label{fig:length_sensitivity}
\end{figure}

\paragraph{Impact of Input Length}
To investigate the capability of modeling long-term dependencies, we conduct a sensitivity analysis on the input sequence length $L$. We vary $L \in \{64, 128, 256, 512, 640\}$ while fixing the prediction horizon at $T=64$. The results, illustrated in Figure~\ref{fig:length_sensitivity}, reveal distinct performance trends. First, regarding \textbf{our model}, the prediction error decreases steadily as the input length extends, confirming that the Learnable Cascaded Wavelet Path effectively leverages extended historical information. Second, regarding \textbf{baselines}, performance gains tend to saturate around $L=512$, followed by slight degradation at $L=640$. Consequently, we standardize the input length to \textbf{$L=512$} for all main experiments to ensure a rigorous comparison where baselines operate at their peak performance.

\subsection{Ablation Study}
\begin{table*}[h]
    \centering
    \caption{\textbf{Detailed Ablation Study.} We report the RMSE and MAE for all variants across all datasets and prediction horizons. \textbf{SWIFT} denotes the full model. \textbf{w/o} stands for without. \textbf{w/o LC-Wavelet} (Learnable Cascaded Wavelet), which replaces the learnable wavelet convolution with a standard convolution; \textbf{w/o MIM} (Multivariate Interaction Module), which removes the entire module; \textbf{w/o MIM-spatial}, which removes the spatial stage; and \textbf{w/o MIM-feature}, which removes the feature stage. Avg is averaged from all four prediction lengths. \best{Red} indicates the best result.}
    \label{ablation-append}
    
    \renewcommand{\arraystretch}{1.15} 
    \setlength{\tabcolsep}{3pt}        

    \begin{adjustbox}{width=0.75\linewidth}
    \begin{tabular}{cc|cc|cc|cc|cc|cc}
        \toprule
        \multicolumn{2}{c|}{Variants} & 
        \multicolumn{2}{c|}{\textbf{SWIFT}} & 
        \multicolumn{2}{c}{w/o MIM} & 
        \multicolumn{2}{c}{w/o MIM-Spatial} & 
        \multicolumn{2}{c}{w/o MIM-Feature} & 
        \multicolumn{2}{c}{w/o LC-Wavelet} \\
        
        \cmidrule(r){1-2} \cmidrule(lr){3-4} \cmidrule(lr){5-6} \cmidrule(lr){7-8} \cmidrule(lr){9-10} \cmidrule(l){11-12}
        \multicolumn{2}{c|}{Metric} & RMSE & MAE & RMSE & MAE & RMSE & MAE & RMSE & MAE & RMSE & MAE \\
        \midrule

        \multirow{5}{*}{\rotatebox{90}{Ali2021}}
        & 64  & \best{0.0989} & \best{0.0695} & 0.1123 & 0.0823 & 0.1103 & 0.0813 & 0.1098 & 0.0796 & 0.1289 & 0.0885 \\
        & 128  & \best{0.1042} & \best{0.0766} & 0.1217 & 0.0833 & 0.1195 & 0.0837 & 0.1144 & 0.0825 & 0.1381 & 0.0894 \\
        & 192  & \best{0.1122} & \best{0.0835} & 0.1259 & 0.0859 & 0.1210 & 0.0867 & 0.1191 & 0.0844 & 0.1434 & 0.0934 \\
        & 256  & \best{0.1184} & \best{0.0875} & 0.1326 & 0.0905 & 0.1245 & 0.0881 & 0.1205 & 0.0882 & 0.1498 & 0.0976 \\
        & Avg & \best{0.1084} & \best{0.0793} & 0.1231 & 0.0855 & 0.1188 & 0.0850 & 0.1159 & 0.0837 & 0.1400 & 0.0922 \\
        \midrule

        \multirow{5}{*}{\rotatebox{90}{Ali2022}}
        & 64  & \best{0.0462} & \best{0.0307} & 0.0476 & 0.0320 & 0.0474 & 0.0319 & 0.0475 & 0.0319 & 0.0486 & 0.0329 \\
        & 128 & \best{0.0475} & \best{0.0318} & 0.0490 & 0.0332 & 0.0487 & 0.0331 & 0.0488 & 0.0330 & 0.0498 & 0.0340 \\
        & 192 & \best{0.0486} & \best{0.0326} & 0.0500 & 0.0340 & 0.0498 & 0.0334 & 0.0494 & 0.0333 & 0.0508 & 0.0347 \\
        & 256 & \best{0.0492} & \best{0.0333} & 0.0508 & 0.0346 & 0.0505 & 0.0338 & 0.0499 & 0.0337 & 0.0514 & 0.0353 \\
        & Avg & \best{0.0479} & \best{0.0321} & 0.0494 & 0.0334 & 0.0491 & 0.0331 & 0.0489 & 0.0330 & 0.0501 & 0.0342 \\
        \midrule

        \multirow{5}{*}{\rotatebox{90}{Fisher}}
        & 64  & \best{0.0658} & \best{0.0412} & 0.0671 & 0.0425 & 0.0671 & 0.0423 & 0.0668 & 0.0423 & 0.0680 & 0.0443 \\
        & 128 & \best{0.0743} & \best{0.0503} & 0.0760 & 0.0528 & 0.0762 & 0.0514 & 0.0756 & 0.0518 & 0.0786 & 0.0532 \\
        & 192 & \best{0.0763} & \best{0.0524} & 0.0789 & 0.0537 & 0.0779 & 0.0539 & 0.0774 & 0.0531 & 0.0801 & 0.0548 \\
        & 256 & \best{0.0783} & \best{0.0541} & 0.0812 & 0.0549 & 0.0805 & 0.0557 & 0.0807 & 0.0546 & 0.0815 & 0.0563 \\
        & Avg & \best{0.0737} & \best{0.0495} & 0.0758 & 0.0510 & 0.0754 & 0.0508 & 0.0751 & 0.0505 & 0.0771 & 0.0521 \\
        \midrule

        \multirow{5}{*}{\rotatebox{90}{Google}}
        & 64  & \best{0.0588} & \best{0.0373} & 0.0604 & 0.0388 & 0.0598 & 0.0382 & 0.0596 & 0.0385 & 0.0618 & 0.0402 \\
        & 128 & \best{0.0609} & \best{0.0396} & 0.0615 & 0.0403 & 0.0611 & 0.0406 & 0.0610 & 0.0409 & 0.0631 & 0.0427 \\
        & 192 & \best{0.0616} & \best{0.0411} & 0.0626 & 0.0432 & 0.0624 & 0.0429 & 0.0620 & 0.0426 & 0.0646 & 0.0449 \\
        & 256 & \best{0.0625} & \best{0.0422} & 0.0637 & 0.0440 & 0.0637 & 0.0436 & 0.0635 & 0.0437 & 0.0660 & 0.0458 \\
        & Avg & \best{0.0610} & \best{0.0400} & 0.0621 & 0.0416 & 0.0617 & 0.0413 & 0.0616 & 0.0414 & 0.0639 & 0.0434 \\
        
        \bottomrule
    \end{tabular}
    \end{adjustbox}
\end{table*}
\paragraph{Detailed Ablation Study Results.} 
The full ablation results are reported in Table~\ref{ablation-append}. By progressively removing or replacing key components, we validate the contribution of each module. The complete SWIFT model consistently achieves the best performance across all datasets and prediction horizons, confirming the necessity of its architectural design.

\subsection{Efficiency Analysis}

\begin{table*}[h]
    \centering
    \caption{\textbf{Detailed Efficiency Analysis.} We report the inference latency (ms/sample) and memory usage (MB) across all prediction lengths for four datasets.  Avg is averaged from all four prediction lengths.}
    \label{efficiency-append}
    
    \renewcommand{\arraystretch}{1.15} 
    \setlength{\tabcolsep}{3pt}        

    \begin{adjustbox}{width=\linewidth}
    \begin{tabular}{cc|cc|cc|cc|cc|cc|cc|cc}
        \toprule
        \multicolumn{2}{c|}{Models} & 
        \multicolumn{2}{c|}{\textbf{SWIFT}} & 
        \multicolumn{2}{c}{TimesNet} & 
        \multicolumn{2}{c}{TimeXer} & 
        \multicolumn{2}{c}{Fremer} & 
        \multicolumn{2}{c}{WDFormer} & 
        \multicolumn{2}{c}{ModernTCN} & 
        \multicolumn{2}{c}{MICN} \\
        
        \cmidrule(r){1-2} \cmidrule(lr){3-4} \cmidrule(lr){5-6} \cmidrule(lr){7-8} \cmidrule(lr){9-10} \cmidrule(lr){11-12} \cmidrule(lr){13-14} \cmidrule(l){15-16}
        \multicolumn{2}{c|}{Metric} & Latency & Mem & Latency & Mem & Latency & Mem & Latency & Mem & Latency & Mem & Latency & Mem & Latency & Mem \\
        \midrule

        \multirow{5}{*}{\rotatebox{90}{Ali2021}}
        & 64  & 3.35 & 134.88 & 16.44 & 170.01 & 7.94 & 50.92 & 6.55 & 50.64 & 9.29 & 65.45 & 4.55 & 187.46 & 4.48 & 143.68 \\
        & 128  & 3.39 & 135.62 & 16.55 & 180.65 & 8.21 & 62.17 & 6.70 & 62.32 & 9.39 & 77.92 & 4.75 & 188.58 & 4.50 & 154.87 \\
        & 192  & 3.41 & 137.89 & 16.78 & 246.44 & 8.22 & 78.16 & 7.57 & 78.90 & 9.60 & 92.52 & 4.83 & 189.70 & 4.50 & 170.49 \\
        & 256  & 3.51 & 138.86 & 17.67 & 212.62 & 8.22 & 85.58 & 7.86 & 85.64 & 9.64 & 102.30 & 4.84 & 198.18 & 4.51 & 177.49 \\
        & Avg & 3.42 & 136.81 & 16.86 & 202.43 & 8.15 & 69.21 & 7.17 & 69.38 & 9.48 & 84.55 & 4.74 & 190.98 & 4.50 & 161.63 \\
        \midrule

        \multirow{5}{*}{\rotatebox{90}{Ali2022}}
        & 64  & 4.47 & 2065.89 & 18.55 & 1124.58 & 10.57 & 760.54 & 8.66 & 745.73 & 11.60 & 757.89 & 4.85 & 2588.41 & 4.54 & 2294.68 \\
        & 128 & 4.50 & 2141.16 & 18.63 & 1631.34 & 11.58 & 1461.60 & 8.66 & 1442.46 & 11.71 & 1454.10 & 4.87 & 3285.52 & 4.60 & 2991.79 \\
        & 192 & 4.51 & 2456.05 & 19.01 & 2572.06 & 12.08 & 2156.77 & 8.68 & 2134.53 & 11.97 & 2146.53 & 4.88 & 3978.55 & 4.76 & 3687.88 \\
        & 256 & 4.60 & 2563.89 & 19.46 & 3282.20 & 12.26 & 2844.15 & 8.74 & 2818.56 & 12.44 & 2828.69 & 4.88 & 4664.33 & 4.79 & 4371.70 \\
        & Avg & 4.52 & 2306.75 & 18.92 & 2152.55 & 11.62 & 1805.76 & 8.69 & 1785.32 & 11.93 & 1796.80 & 4.87 & 3629.20 & 4.68 & 3336.51 \\
        \midrule

        \multirow{5}{*}{\rotatebox{90}{Fisher}}
        & 64  & 3.81 & 121.44 & 17.70 & 304.54 & 10.89 & 56.70 & 7.42 & 52.37 & 11.46 & 67.84 & 3.85 & 216.24 & 4.18 & 185.88 \\
        & 128 & 3.85 & 145.20 & 17.73 & 344.20 & 11.63 & 80.56 & 7.47 & 73.71 & 11.72 & 88.67 & 3.94 & 240.70 & 4.21 & 188.10 \\
        & 192 & 3.86 & 166.21 & 18.08 & 533.64 & 11.82 & 104.20 & 7.48 & 93.82 & 11.77 & 108.27 & 3.94 & 262.43 & 4.26 & 190.58 \\
        & 256 & 3.87 & 188.29 & 18.31 & 578.46 & 12.02 & 128.09 & 7.56 & 114.51 & 12.19 & 128.45 & 3.97 & 284.97 & 4.29 & 192.13 \\
        & Avg & 3.85 & 155.28 & 17.96 & 440.21 & 11.59 & 92.39 & 7.48 & 83.60 & 11.78 & 98.31 & 3.93 & 251.09 & 4.24 & 189.17 \\
        \midrule

        \multirow{5}{*}{\rotatebox{90}{Google}}
        & 64  & 3.68 & 130.74 & 17.15 & 314.19 & 10.79 & 66.14 & 7.42 & 61.58 & 10.05 & 77.05 & 3.70 & 226.61 & 4.31 & 186.39 \\
        & 128 & 3.69 & 161.73 & 17.32 & 361.19 & 10.80 & 99.41 & 7.44 & 92.43 & 10.46 & 107.39 & 3.71 & 256.85 & 4.47 & 189.65 \\
        & 192 & 3.73 & 194.44 & 17.37 & 406.27 & 10.87 & 132.24 & 7.44 & 121.87 & 11.40 & 136.32 & 3.73 & 289.59 & 4.47 & 193.18 \\
        & 256 & 3.77 & 225.87 & 17.47 & 616.03 & 10.93 & 164.14 & 7.54 & 150.56 & 11.40 & 164.50 & 3.76 & 322.59 & 4.48 & 196.92 \\
        & Avg & 3.72 & 178.19 & 17.32 & 424.42 & 10.85 & 115.48 & 7.46 & 106.61 & 10.83 & 121.31 & 3.72 & 273.91 & 4.43 & 191.53 \\
        
        \bottomrule
    \end{tabular}
    \end{adjustbox}
\end{table*}

\paragraph{Detailed Efficiency Analysis.} 
The comprehensive efficiency comparison, including inference latency and memory usage across all prediction horizons, is detailed in Table~\ref{efficiency-append}. SWIFT demonstrates superior computational efficiency, achieving the lowest average latency across all datasets. Regarding memory consumption, while SWIFT exhibits a moderate increase on the high-dimensional Ali2022 dataset due to its expanded channel modeling, it remains highly efficient on standard benchmarks (Ali2021, Fisher, Google), striking an optimal balance between high forecasting accuracy and low resource overhead.

\begin{table*}[h]
    \centering
    \caption{\textbf{Universality Verification on ETT Benchmarks}. To verify the universality of SWIFT beyond workload prediction, we evaluate it on ETT benchmarks. We report the full performance breakdown across four prediction horizons $T \in \{96, 192, 336, 720\}$. Avg is averaged from all four prediction lengths. \best{Red} indicates the best result.}
    \label{long-term-ett-append}
    
    \renewcommand{\arraystretch}{1} 
    \setlength{\tabcolsep}{3pt}        

    \begin{adjustbox}{width=0.9\linewidth}
    \begin{tabular}{cc|cc|cc|cc|cc|cc|cc|cc}
        \toprule
        \multicolumn{2}{c|}{Models} & \multicolumn{2}{c|}{\textbf{SWIFT}} & \multicolumn{2}{c}{TimesNet} & \multicolumn{2}{c}{TimeXer} & \multicolumn{2}{c}{Fremer} & \multicolumn{2}{c}{WDFormer} & \multicolumn{2}{c}{ModernTCN} & \multicolumn{2}{c}{MICN} \\ 
        
        \cmidrule(r){1-2} \cmidrule(lr){3-4} \cmidrule(lr){5-6} \cmidrule(lr){7-8} \cmidrule(lr){9-10} \cmidrule(lr){11-12} \cmidrule(lr){13-14} \cmidrule(l){15-16}
        \multicolumn{2}{c|}{Metric} & MSE & MAE & MSE & MAE & MSE & MAE & MSE & MAE & MSE & MAE & MSE & MAE & MSE & MAE \\
        \midrule
        \multirow{4}{*}{\rotatebox{90}{ETTh1}}
        & 96 & \best{0.356} & \best{0.388} & 0.422 & 0.424 & 0.390 & 0.400 & 0.384 & 0.395 & 0.388 & 0.405 & 0.377 & 0.397 & 0.404 & 0.402 \\
        & 192 & \best{0.430} & \best{0.419} & 0.483 & 0.460 & 0.440 & 0.429 & 0.436 & 0.428 & 0.431 & 0.430 & 0.440 & 0.430 & 0.470 & 0.443 \\
        & 336 & \best{0.466} & \best{0.441} & 0.512 & 0.479 & 0.479 & 0.448 & 0.478 & 0.450 & 0.471 & 0.452 & 0.477 & 0.451 & 0.498 & 0.476 \\
        & 720 & \best{0.463} & \best{0.463} & 0.535 & 0.520 & 0.468 & 0.463 & 0.469 & 0.465 & 0.464 & 0.468 & 0.489 & 0.493 & 0.524 & 0.518 \\
        & Avg & \best{0.429} & \best{0.428} & 0.488 & 0.471 & 0.444 & 0.435 & 0.442 & 0.435 & 0.439 & 0.439 & 0.446 & 0.443 & 0.474 & 0.460 \\
        \midrule
        \multirow{4}{*}{\rotatebox{90}{ETTh2}}
        & 96 & \best{0.285} & \best{0.334} & 0.315 & 0.360 & 0.287 & 0.338 & 0.301 & 0.347 & 0.302 & 0.353 & 0.301 & 0.347 & 0.316 & 0.380 \\
        & 192 & \best{0.373} & 0.394 & 0.404 & 0.415 & 0.376 & \best{0.391} & 0.385 & 0.396 & 0.384 & 0.403 & 0.386 & 0.398 & 0.390 & 0.418 \\
        & 336 & 0.427 & \best{0.431} & 0.448 & 0.447 & \best{0.425} & 0.432 & 0.432 & 0.435 & 0.430 & 0.437 & 0.439 & 0.441 & 0.442 & 0.461 \\
        & 720 & 0.442 & 0.448 & 0.467 & 0.468 & 0.442 & 0.447 & 0.430 & \best{0.446} & \best{0.428} & 0.446 & 0.461 & 0.461 & 0.468 & 0.475 \\
        & Avg & \best{0.381} & \best{0.401} & 0.409 & 0.423 & 0.382 & 0.402 & 0.387 & 0.406 & 0.386 & 0.410 & 0.397 & 0.412 & 0.404 & 0.433 \\
        \midrule
        \multirow{4}{*}{\rotatebox{90}{ETTm1}}
        & 96 & \best{0.485} & 0.490 & 0.524 & 0.518 & 0.519 & 0.523 & 0.572 & 0.553 & 0.495 & 0.502 & 0.490 & \best{0.488} & 0.511 & 0.518 \\
        & 192 & 0.590 & \best{0.533} & 0.630 & 0.561 & 0.611 & 0.550 & 0.627 & 0.565 & \best{0.588} & 0.545 & 0.593 & 0.536 & 0.627 & 0.587 \\
        & 336 & \best{0.726} & \best{0.587} & 0.755 & 0.623 & 0.764 & 0.607 & 0.750 & 0.608 & 0.736 & 0.599 & 0.729 & 0.591 & 0.768 & 0.627 \\
        & 720 & 1.114 & \best{0.694} & 1.181 & 0.736 & 1.143 & 0.705 & \best{1.095} & 0.696 & 1.147 & 0.709 & 1.123 & 0.696 & 1.192 & 0.750 \\
        & Avg & \best{0.729} & \best{0.576} & 0.772 & 0.609 & 0.759 & 0.596 & 0.761 & 0.605 & 0.741 & 0.589 & 0.734 & 0.578 & 0.774 & 0.620 \\
        \midrule
        \multirow{4}{*}{\rotatebox{90}{ETTm2}}
        & 96 & \best{0.653} & \best{0.549} & 0.674 & 0.572 & 0.684 & 0.571 & 0.700 & 0.582 & 0.672 & 0.567 & 0.672 & 0.556 & 0.688 & 0.591 \\
        & 192 & \best{0.818} & \best{0.629} & 0.839 & 0.652 & 0.841 & 0.646 & 0.859 & 0.654 & 0.835 & 0.646 & 0.828 & 0.633 & 0.842 & 0.668 \\
        & 336 & \best{1.098} & \best{0.744} & 1.127 & 0.762 & 1.134 & 0.760 & 1.148 & 0.765 & 1.128 & 0.761 & 1.112 & 0.744 & 1.161 & 0.774 \\
        & 720 & \best{1.495} & 0.876 & 1.539 & 0.891 & 1.531 & 0.891 & 1.565 & 0.897 & 1.504 & 0.885 & 1.503 & \best{0.875} & 1.603 & 0.904 \\
        & Avg & \best{1.016} & \best{0.700} & 1.045 & 0.719 & 1.048 & 0.717 & 1.068 & 0.724 & 1.035 & 0.715 & 1.029 & 0.702 & 1.073 & 0.734 \\
        \midrule
        \multicolumn{2}{c|}{1st Count} & \best{16} & \best{16} & 0 & 0 & 1 & 1 & 1 & 1 & 2 & 0 & 0 & 2 & 0 & 0 \\
        \bottomrule
    \end{tabular}
    \end{adjustbox}
\end{table*}

\begin{table*}[h]
    \centering
    \caption{\textbf{Stability Analysis.} We report the standard deviation of RMSE ($\sigma_{RMSE}$) and MAE ($\sigma_{MAE}$) across all four prediction horizons $T \in \{64, 128, 192, 256\}$.}
    \label{tab:stability_analysis}
    
    \renewcommand{\arraystretch}{1.2} 
    \setlength{\tabcolsep}{4pt}       

    \begin{adjustbox}{width=\linewidth}
    \begin{tabular}{cc|cc|cc|cc|cc|cc|cc|cc}
        \toprule
        \multicolumn{2}{c|}{Models} & 
        \multicolumn{2}{c|}{\textbf{SWIFT}} & 
        \multicolumn{2}{c}{TimesNet} &   
        \multicolumn{2}{c}{TimeXer} &       
        \multicolumn{2}{c}{Fremer} &       
        \multicolumn{2}{c}{WDFormer} &        
        \multicolumn{2}{c}{ModernTCN} &    
        \multicolumn{2}{c}{MICN} \\    
        
        \cmidrule(r){1-2} \cmidrule(lr){3-4} \cmidrule(lr){5-6} \cmidrule(lr){7-8} \cmidrule(lr){9-10} \cmidrule(lr){11-12} \cmidrule(lr){13-14} \cmidrule(l){15-16}
        \multicolumn{2}{c|}{Metric} & $\sigma_{RMSE}$ & $\sigma_{MAE}$ & $\sigma_{RMSE}$ & $\sigma_{MAE}$ & $\sigma_{RMSE}$ & $\sigma_{MAE}$ & $\sigma_{RMSE}$ & $\sigma_{MAE}$ & $\sigma_{RMSE}$ & $\sigma_{MAE}$ & $\sigma_{RMSE}$ & $\sigma_{MAE}$ & $\sigma_{RMSE}$ & $\sigma_{MAE}$ \\
        \midrule

        \multirow{4}{*}{\rotatebox{90}{Ali2021}}
        & 64  & 0.0016 & 0.0014 & 0.0029 & 0.0034 & 0.0049 & 0.0023 & 0.0014 & 0.0013 & 0.0045 & 0.0040 & 0.0015 & 0.0013 & 0.0025 & 0.0020 \\
        & 128  & 0.0010 & 0.0018 & 0.0044 & 0.0038 & 0.0025 & 0.0032 & 0.0010 & 0.0016 & 0.0043 & 0.0031 & 0.0011 & 0.0021 & 0.0040 & 0.0036 \\
        & 192  & 0.0014 & 0.0016 & 0.0031 & 0.0020 & 0.0035 & 0.0033 & 0.0013 & 0.0014 & 0.0039 & 0.0024 & 0.0018 & 0.0021 & 0.0032 & 0.0024 \\
        & 256  & 0.0025 & 0.0015 & 0.0034 & 0.0039 & 0.0042 & 0.0035 & 0.0026 & 0.0019 & 0.0035 & 0.0031 & 0.0024 & 0.0012 & 0.0032 & 0.0034 \\
        \midrule

        \multirow{4}{*}{\rotatebox{90}{Ali2022}}
        & 64  & 0.0015 & 0.0016 & 0.0049 & 0.0034 & 0.0046 & 0.0028 & 0.0017 & 0.0019 & 0.0035 & 0.0038 & 0.0017 & 0.0021 & 0.0037 & 0.0026 \\
        & 128  & 0.0025 & 0.0018 & 0.0037 & 0.0023 & 0.0035 & 0.0034 & 0.0031 & 0.0020 & 0.0048 & 0.0034 & 0.0028 & 0.0022 & 0.0049 & 0.0038 \\
        & 192  & 0.0016 & 0.0017 & 0.0048 & 0.0040 & 0.0040 & 0.0029 & 0.0018 & 0.0019 & 0.0046 & 0.0025 & 0.0014 & 0.0015 & 0.0036 & 0.0024 \\
        & 256  & 0.0022 & 0.0018 & 0.0036 & 0.0037 & 0.0048 & 0.0030 & 0.0024 & 0.0023 & 0.0025 & 0.0027 & 0.0024 & 0.0022 & 0.0044 & 0.0024 \\
        \midrule

        \multirow{4}{*}{\rotatebox{90}{Fisher}}
        & 64  & 0.0012 & 0.0011 & 0.0046 & 0.0020 & 0.0050 & 0.0033 & 0.0014 & 0.0014 & 0.0043 & 0.0038 & 0.0013 & 0.0013 & 0.0025 & 0.0024 \\
        & 128  & 0.0022 & 0.0015 & 0.0046 & 0.0039 & 0.0049 & 0.0025 & 0.0023 & 0.0016 & 0.0027 & 0.0020 & 0.0019 & 0.0014 & 0.0038 & 0.0031 \\
        & 192  & 0.0011 & 0.0012 & 0.0038 & 0.0023 & 0.0036 & 0.0036 & 0.0012 & 0.0014 & 0.0046 & 0.0038 & 0.0010 & 0.0011 & 0.0037 & 0.0028 \\
        & 256  & 0.0013 & 0.0015 & 0.0049 & 0.0028 & 0.0048 & 0.0040 & 0.0011 & 0.0015 & 0.0037 & 0.0023 & 0.0014 & 0.0016 & 0.0046 & 0.0028 \\
        \midrule

        \multirow{4}{*}{\rotatebox{90}{Google}}
        & 64  & 0.0011 & 0.0010 & 0.0031 & 0.0035 & 0.0029 & 0.0037 & 0.0012 & 0.0013 & 0.0030 & 0.0035 & 0.0010 & 0.0008 & 0.0038 & 0.0029 \\
        & 128  & 0.0016 & 0.0011 & 0.0027 & 0.0025 & 0.0037 & 0.0032 & 0.0018 & 0.0014 & 0.0039 & 0.0027 & 0.0021 & 0.0013 & 0.0027 & 0.0037 \\
        & 192  & 0.0018 & 0.0017 & 0.0034 & 0.0036 & 0.0030 & 0.0025 & 0.0021 & 0.0021 & 0.0048 & 0.0036 & 0.0023 & 0.0020 & 0.0037 & 0.0025 \\
        & 256  & 0.0017 & 0.0010 & 0.0048 & 0.0040 & 0.0041 & 0.0025 & 0.0016 & 0.0009 & 0.0043 & 0.0026 & 0.0020 & 0.0012 & 0.0031 & 0.0023 \\
        \midrule
    \end{tabular}
    \end{adjustbox}
\end{table*}

\subsection{Universality Verification}

\paragraph{Detailed Universality Verification Analysis.}
To verify the universality of our model beyond the specific domain of workload prediction, we conducted extensive experiments on the standard ETT benchmarks. As presented in Table~\ref{long-term-ett-append}, SWIFT demonstrates exceptional universality, achieving sota performance in \textbf{32 out of 40} cases. Quantitatively, SWIFT outperforms the Transformer-based model with a maximum error reduction of \textbf{6.12\%}. Compared to the CNN-based baseline, SWIFT achieves a significant improvement of up to \textbf{15.56\%}. These results confirm that SWIFT is not limited to server workload patterns but effectively capture the underlying physical dynamics and complex temporal correlations in general time-series data.

\subsection{Stability Analysis}

\paragraph{Stability Analysis.} 
The stability of the proposed SWIFT is evaluated by measuring the standard deviation of RMSE and MAE across three independent runs with different random seeds, as shown in Table~\ref{tab:stability_analysis}. Overall, SWIFT demonstrates robust stability, maintaining consistently low standard deviations across most datasets and prediction horizons. This indicates that our model is insensitive to initialization and training randomness, showcasing its great stability and reliability for practical deployment.

\clearpage
\section{Showcases}
\begin{figure}[H]
    \centering
    \includegraphics[width=\linewidth]{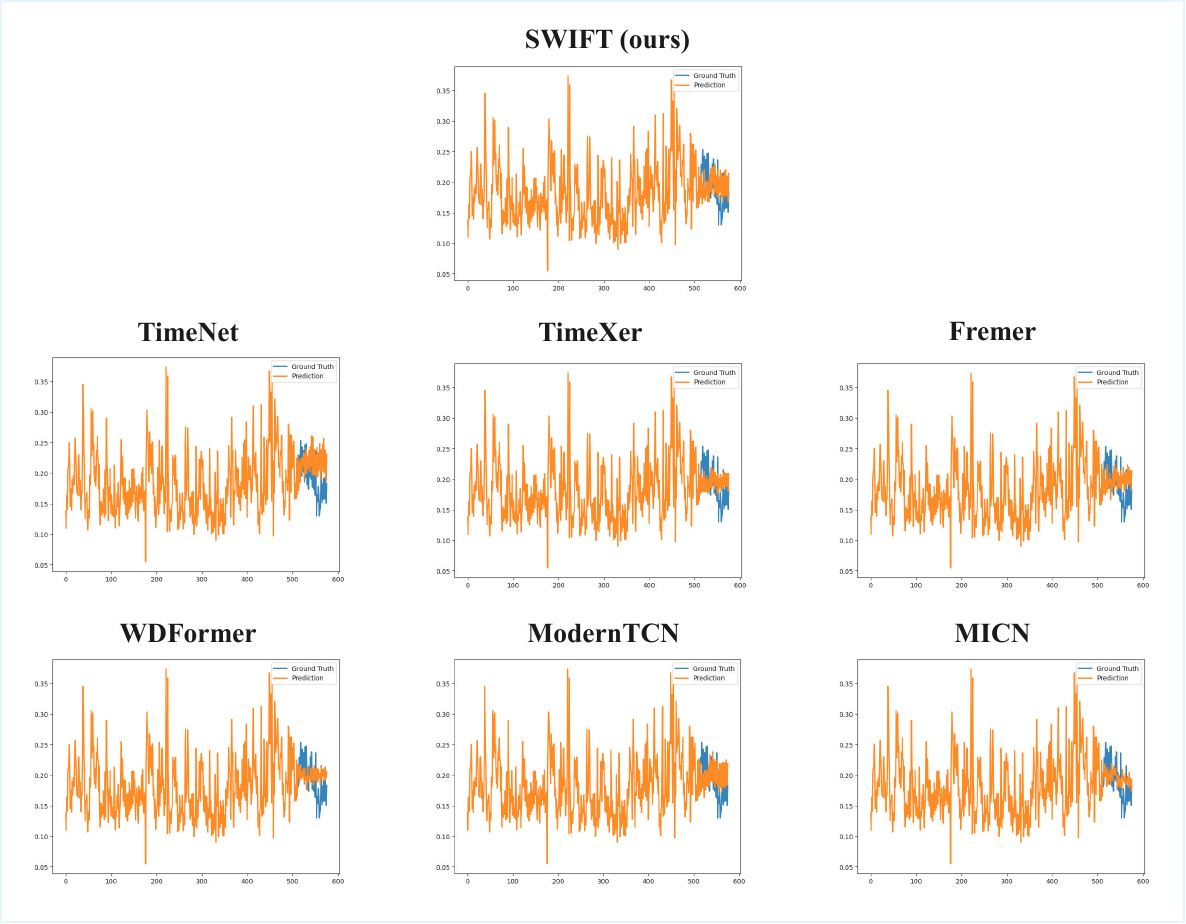} 
    \caption{Visualization of long-term workload forecasting on Google under the input-512-predict-64 setting. The blue lines stand for the ground truth and the orange lines stand for predicted values.}
    \label{fig:google-showcase}
\end{figure}

\begin{figure}[H]
    \centering
    \includegraphics[width=\linewidth]{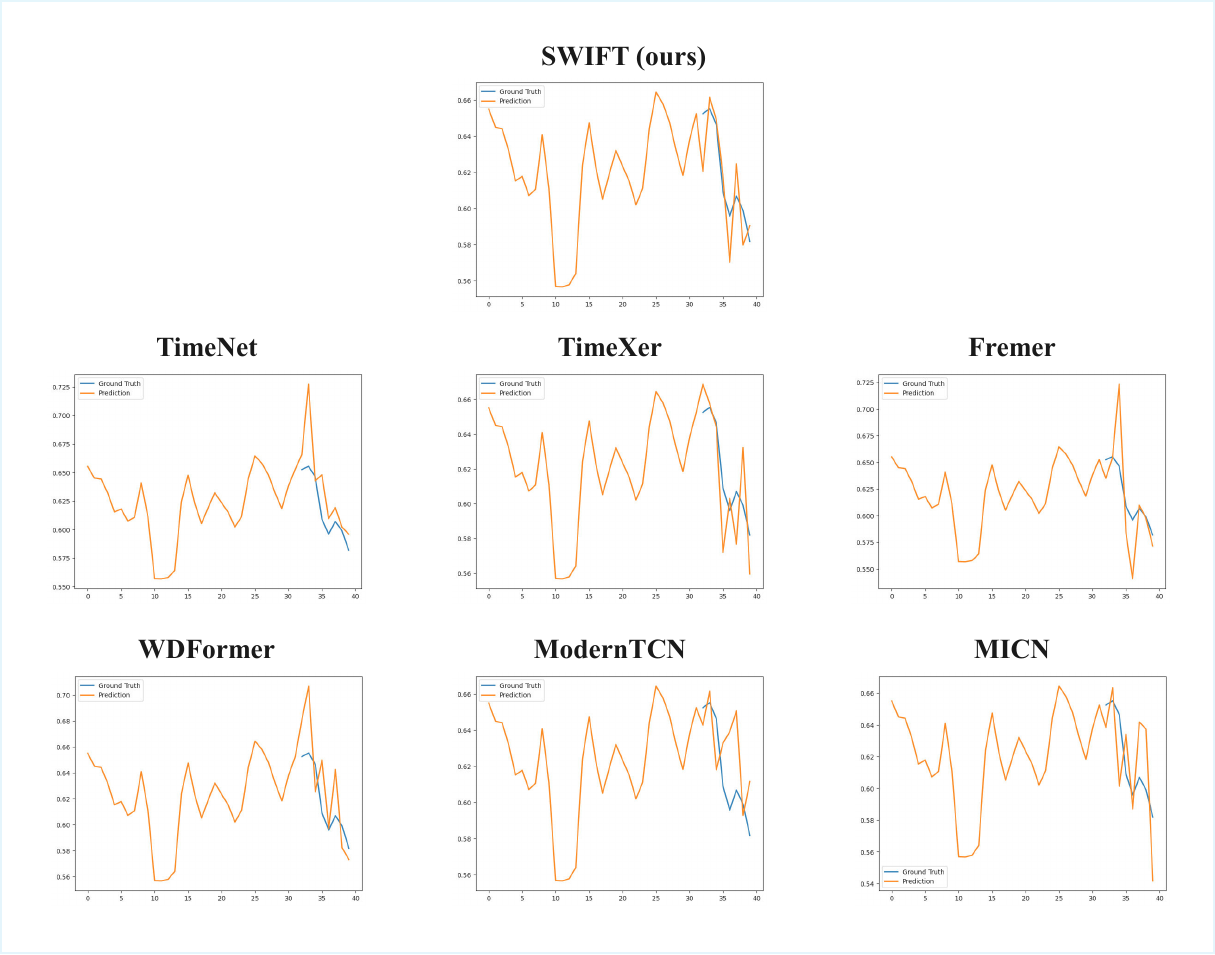}
    \caption{Visualization of short-term workload forecasting on Ali2021 under the input-32-predict-8 setting. The blue lines stand for the ground truth and the orange lines stand for predicted values.}
    \label{fig:alibaba-showcase}
\end{figure}

\begin{figure}[H]
    \centering
    \includegraphics[width=\linewidth]{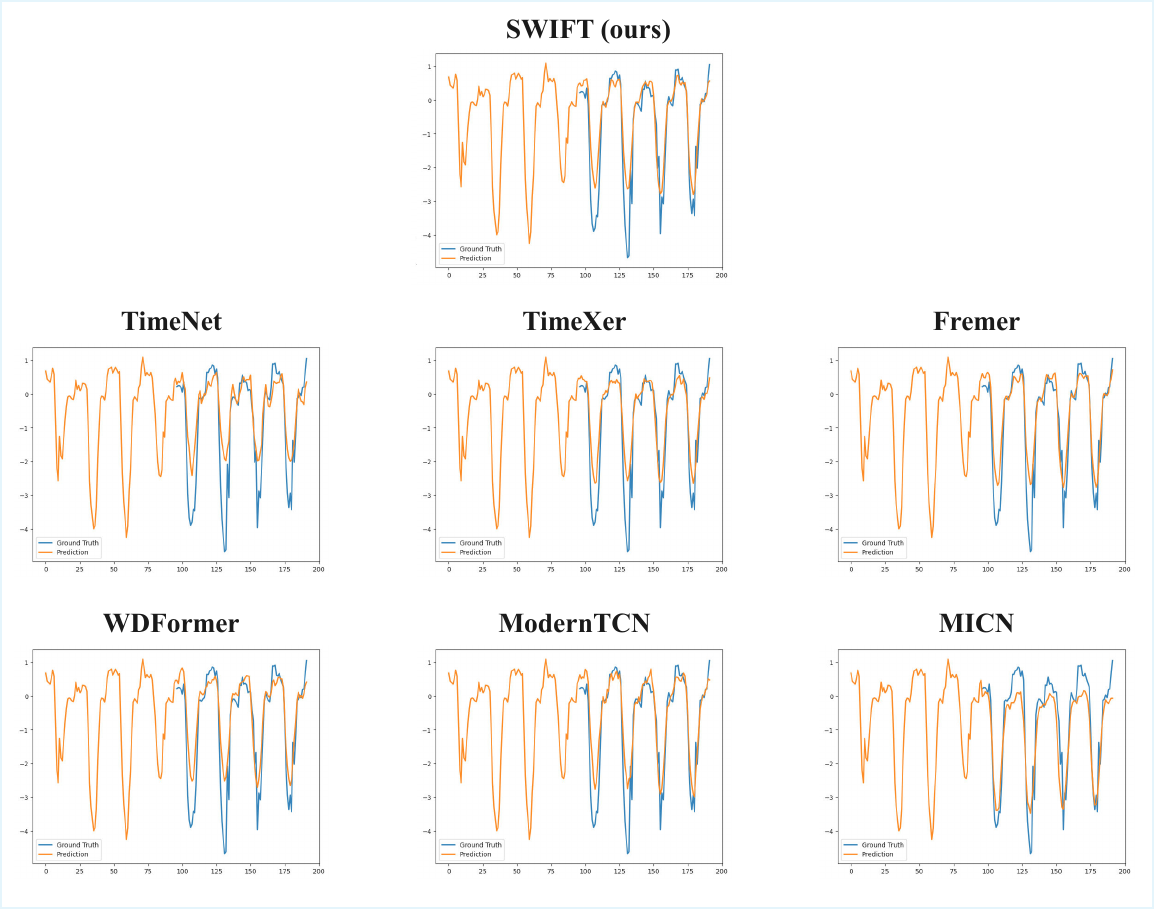}
    \caption{Visualization of general time-series forecasting on ETTh1 under the input-96-predict-96 setting. The blue lines stand for the ground truth and the orange lines stand for predicted values.}
    \label{fig:ett-showcase}
\end{figure}

To provide an intuitive comparison among different models, we present a series of visualization showcases across three distinct scenarios: long-term workload forecasting (Figure~\ref{fig:google-showcase}), short-term workload forecasting(Figure~\ref{fig:alibaba-showcase}), and general time-series forecasting(Figure~\ref{fig:ett-showcase}).
\subsection{Long-Term Workload Forecasting}

\clearpage
\subsection{Short-Term Workload Forecasting}

\clearpage
\subsection{General Time-Series Forecasting}

\end{document}